\documentclass{ws-ijmpd}
\usepackage[super,compress]{cite}
\begin{document}

\markboth{Sohrab Rahvar}
{Gravitational Microlensing I: A Unique Astrophysical Tool}

%
\catchline{}{}{}{}{}
%

\title{Gravitational Microlensing I: A Unique Astrophysical Tool 
}

\author{Sohrab Rahvar
}

\address{Department of Physics, Sharif University of
Technology,\\
 P.O.Box 11365--9161, Tehran, Iran\\
rahvar@sharif.edu}



\maketitle

\begin{history}
\received{Day Month Year}
\revised{Day Month Year}
\end{history}

\begin{abstract}
In this article we review the astrophysical application of gravitational microlensing. After  introducing the history of gravitational lensing, we present the key equations and concept of microlensing. The most frequent microlensing events are single-lens events and historically it has been used for searching dark matter in the form of compact astrophysical halo objects in the Galactic halo. We discuss about the degeneracy problem in the parameters of lens and perturbation effects that can partially break the degeneracy between the lens parameters. The rest of paper is about the astrophysical applications of microlensing. One of the important applications is in  the stellar physics by probing the surface of source stars in the high magnification microlensing events. The astrometric and polarimetric observations will be 
complimentary for probing the atmosphere and stellar spots on the surface of source stars.  
Finally we discuss about the future projects as space based telescopes for parallax and astrometry observations of microlensing events. With this project, we would expect to produce a complete stellar and remnant mass function and study the structure of Galaxy in term of distribution of stars along our line of sight towards the centre of galaxy. 
 
\end{abstract}

\keywords{Gravitational Microlensing; MACHOs; Stellar Physics}

\ccode{PACS numbers:}



 
  
 

\section{Introduction}
In this article we will review the astrophysical applications of a specific type of gravitational lensing, so-called  gravitational microlensing, in the Milky Way stars. The bending of light in classical mechanics, when a light ray passes by a massive object assuming a mass for the photons is foreseen. The extension of this idea to the curved space-time in general relativity leads us to a correct value for the light bending which is compatible with the observations. 

While in the gravitational lensing more than one image is produced, in microlensing due to small separation of images, only the magnification effect of a transit lens over a background star can be measured. The application of microlensing for exploring the dark matter of Galactic halo answered to one of main astrophysical questions about the nature of dark matter in the halo of galaxy.  After one decade monitoring of Large and Small Magellanic Clouds by three main observational groups, they conclude that dark matter in the halo of Milky Way is not made of Massive Astrophysical Compact Halo Objects (MACHOs).

The other application of microlensing is the possibility of detecting extrasolar planets. A planet or a group of planets that orbit around the parent star can play the role of multiple lensing in the microlensing events. In this case, parent star with the planets around it behave as double or multiple lenses and the result is producing 
caustic lines on the source plane. Once a source star crosses these singular lines, a small 
perturbation on the light curve appears and result is short blip or dip in the light curve. Analyzing the light curve reveals characteristics such as distance of planet from the parent star and mass of an exoplanet \cite{gplanet,gaudi}. Gravitational microlensing compare to the other methods of planet observations, is a unique tool to detect planets beyond the snow line as well as planets orbiting in the habitable zone of the parent stars \cite{habit,planet1,planet2,planet3,planet4,planet5,planet6,planet7,planet9,planet10}. The detection of exoplanets by microlensing is a well established method and next generation of microlensing experiments with high cadence and better photometric precision will detect a larger number of exoplanets \cite{KMnet}.
 
The new area of gravitational microlensing experiments would be the astrophysical application of this phenomenon that we will review the details of this subject in this paper. The gravitational microlensing is a natural telescope that can probe the surface of source stars, located few thousands light years far from us \cite{pascal}. The surface of stars are not uniform and due to scattering of light from the atmosphere, for an observer, it seems darker at the edge of star and brighter at the central part of it. Moreover, stellar spots are defects on the surface of star by means of cooler areas compare to the background of star. In the case of high magnification microlensing events, lens probes surface of source star and it can encode the surface brightness of star in the microlensing light curve. 

The other possibility in the microlensing experiment is the astrometry of images as well as polarimetry observations. Both methods will provide extra information beside the photometric observations. In the case of astrometry which can be carried out by the space-based telescopes or high resolution ground based telescopes, we can measure the time variation of centre of light for the source stars during microlensing. Astrometric observations directly provides the size of Einstein angle. On the other hand polarimetric observations can identify the local polarization on the surface of source star and also identify the spots on the stars. This observation will open a new window to the stellar studies of remote stars and investigate the physics of stars with more details. The other advantage of a space based telescope orbiting around Sun is that we can make parallax measurements and with the extra information from the astrometry observations, it is possible to identify the mass and distance of lenses from the earth. This would be a great possibility for measuring the mass function of visible and stellar remnants, as well as the distribution of stars in the Milky Way.

In section (\ref{history}) we will provide a brief history of gravitational lensing and microlensing. In section (\ref{formalism}), formalism of gravitational lensing, including the lens equation is discussed. Section (\ref{statistics}) introduces calculation for the statistics of microlensing events in different directions of Milky-Way galaxy. In section (\ref{perturbation}), we will discuss about the perturbation effects as the parallax and finite-size effects on the light curve of microlensing events and applications of these effects for breaking the degeneracy between the lens parameters. Also small perturbations on the light curve by the effects of stellar spots and compact objects in binary systems will be discussed in this section. In section (\ref{astrometry}), we review the astrometry and polarimetry observations of microlensing events and our prospects for future observations and astrophysical applications of polarimetry.  The summary is given in section (\ref{conc}).

\section{A brief history of Gravitational Lensing}
\label{history}
 In Newtonian gravity, assuming photons are made of tiny particles and moving with the speed of 
 light, one can predict the bending of light bundle while passing beside a source of gravity. We will use the term of gravitational lens for the gravitational deflector of light. For a light ray travelling from a distant source to the observer, let us take the minimum distance of the light ray to a point mass lens to be $"\xi"$. Then, the gravitational acceleration perpendicular 
 to the propagation of light ray is given by
 $$ a_y = - \frac{GM}{\xi^2 + x^2}\times \frac{\xi}{\sqrt{\xi^2 + x^2}},$$ where $x = ct$ and the time is set to start from $-\infty$ and goes to $+\infty$. We can integrate acceleration in the $y$-direction and calculate the velocity of 
 a test particle in this direction. The bending angle at each point along the trajectory of light is simply giving by $\alpha = v_y/v_x$ and the overall bending angle is  
 \begin{equation}
 \alpha =   \frac{2 GM}{\xi c^2}.
 \label{Nbending}
 \end{equation}

 In general relativity which is the extension of special relativity to the accelerating frames, the light bending is calculated by the deviation of trajectory of the light rays as a null geodesics in the space-time. The standard approach is using Schwarzschild metric
 \begin{equation}
 ds^2 = -dt^2(1-\frac{2GM}{r}) + dr^2(1-\frac{2GM}{r})^{-1} + r^2 d\theta^2 + r^2\sin^2\theta d\phi^2, 
 \end{equation} 
which describes the metric of space-time around a point mass object.
We can investigate the bending of light rays through Fermat 
principle in which light rays chooses the extremum path for traveling from one point to another point. In the case of using general relativity, the bending angle is twice of that in the Newtonian gravity, given in equation (\ref{Nbending}).  The prediction for bending of light in General Relativity has been confirm by Eddington in 1919. A full story of expedition for 1919 eclipse can be found 
in Ref. \cite{ee}.

The observability of gravitational lensing where one star plays the role of lensing of another background
star has been motivated to Einstein by a Czech engineer, Rudi W. Mandl in 1936 and Einstein wrote a brief paper 
by his request in Science \cite{Ein1936}. However, Einstein noted in his paper " that is unlike to 
detect this phenomenon". One year later a Swiss astronomer Fritz Zwicky noted that while the probability of observation of lensing of a star by another star is low, however in cosmological scales it is likely that 
galaxies can lens the light of other background galaxies \cite{zwik}.

In 1986 Bohdan Paczynski returned to the question by Einstein in 1936 whether the gravitational lensing of 
a star by another star inside our galaxy is observable \cite{pac86}. He calculate the observability of this effect and proposed this method for indirect observation of Massive Astrophysical Compact Halo Objects (MACHOs) in the Galactic halo. Assuming that $100\%$ of halo is made of 
MACHOs, the probability of detecting gravitational lensing by monitoring stars in the Large and Small Magellanic Clouds 
is $10^{-7}$. The leading groups of EROS\footnote{Expérience pour la Recherche d'Objets Sombres} and MACHO started 
monitoring Magellanic Clouds as well as the centre of Galaxy for searching microlensing events and the first microlensing candidate is found in 1993 at the direction of Large Magellanic Clouds \cite{m,e}. After a decade of monitoring Magellanic Clouds, these groups compared the expected 
number of microlensing events if $100\%$ of halo is made of MACHOs, with the observational results. By counting the number of observed microlensing events, the observation groups came to the conclusion of 
putting an upper limit of $20\%$ contribution of MACHOs in the Galactic halo \cite{mr,er}.

While the detection of MACHOs was one of the main goals of gravitational microlensing experiments, the application 
of this method for exploring exoplanets and studying the atmosphere of distant stars opened a new window to the astrophysical applications of this new tool. The idea of 
exploring binary lenses by the gravitational lensing has been proposed by Moa and Paczynski in 1991. In the later years this method become an important tool for detecting planets orbiting stars 
few kilo parsec far from us \cite{mp91}. They not only could detected earth like planets, also their distances from the parent stars were about few astronomical unit, possibility of being in the habitable zone. Moreover, due to control on the defection efficiency, this method is able to provide correct statistics from the planet census.

The new surveys as GAIA and space based microlensing telescopes \cite{dong} as well as ground based telescope \cite{KT} for the astrometric observation will enable us in near future to break the degeneracy between the lens parameters and identify the parameters of lenses. 
One of the main achievement will be the full determination of mass function of all compact objects from the brown dwarf to the blackholes. The polarimetry observations also will reveal the atmospheric structure of a source star as well as the stellar spots of stars. 


 
\section{Formalism of Gravitational Microlensing}
\label{formalism}
The modern area of microlensing started with the paper of Paczynski in 1986 who proposed the observation of lensing of a source star by another star in the Milky Way galaxy. In this type of lensing, the separation between the images from lensing is so small to be resolved by the ground based telescopes. However, the magnification of the background star during lensing is observable. In gravitational lensing inside the Milky Way galaxy, unlike to the cosmological lensing, the positions of observer, lens and source 
are moving with respect to each other and result is a transient lensing effect, so-called gravitation microlensing. In this type of gravitational lensing, the lens magnifies the background star in a time scale of one month. 

\begin{figure}
\centerline{\psfig{file=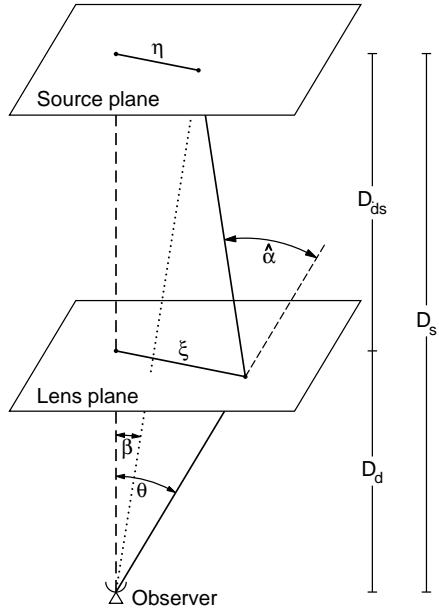,width=6.0cm}}
\vspace*{1pt}
\caption{A schematic illustration of gravitational lensing with the observer, lens and source. This figure is adapted from Ref. \cite{f1} \label{fig1}}
\end{figure}

For the gravitational lensing formalism, we use the simple geometry as shown in Figure (\ref{fig1}). Here the deflection angle is given by $\hat{\alpha}$ and it is produced by a single lens but we can also extend it to the multiple lenses . Using simple 
geometrical relation, we can write the relation between the 
position of unlensed source $\beta$ and images after lensing $\theta$, as follows:
\begin{equation}
{\hat\theta} D_s = {\hat\beta} D_s + {\hat\alpha} D_{ds},
\label{leq}
\end{equation}
where the hat sign represents angle as a two dimensional vector perpendicular to our line of sight, $D_s$ is the distance of observer to the source, $D_{d}$ is the distance of observer to the deflector (lens) and $D_{ds}$ is the distance of deflector to the source. For the single lens, the deflection angle is given by
\begin{equation}
\hat{\alpha} = \frac{4 G M}{c^2 D_d}\frac{\hat\theta}{\theta^2},
\label{dif}
\end{equation}
where it can be written as $|\hat{\alpha}| = 4\Phi/c^2$ and $\Phi$ is the Newtonian potential.

Since in the linear theory of General Relativity, the perturbation terms have linear property, for the case of having multiple lenses at the same distances from the observer, the overall 
deflection angle can be written as the superposition of deflection angles from each lens
\begin{equation}
\hat{\alpha} = \sum_i\frac{4 G M_i}{c^2 D_d}\frac{\hat\theta-\hat\theta_i}{|\hat\theta-\hat\theta_i|^2}.
\end{equation}
For the case of single lens, substituting equation (\ref{dif}) in equation (\ref{leq}), we can write the following equation that relates the position of source to the position of lens:
\begin{equation}
\theta^2 = \beta\theta + \theta_E^2,
\label{le2}
\end{equation}
where the Einstein angle is define by 
\begin{equation}
\theta_E^2 = \frac{4GM}{c^2} \frac{D_{ds}}{D_dD_s}.
\label{tang}
\end{equation}
By solving equation (\ref{le2}) for single lens, we find two solutions for this equation 
as follows:
\begin{equation}
\theta = \frac{\beta\pm\sqrt{\beta^2 + 4\theta_E^2}}{2}.
\label{sol1}
\end{equation}
In the single lensing, the mapping is radial on the lens plane and the distortion in the images results 
from this radial transformation. The flux of stars on the lens plane is approximately constant and it is independent of  the position of the lens plane \cite{rahvar2014}. Hence, having larger images from the lensing compare to the unlensed source star results in receiving more flux from the source star. The amount of magnification is the ratio of the image area to the source area as follows
\begin{equation}
A = |\frac{\theta d\theta}{\beta d\beta}|.
\end{equation}
From equation (\ref{sol1}), we have two solutions for this equation, then the total magnification would be the sum of magnifications by two images as 
\begin{eqnarray}
A &=& A^+ +  A^- \\
A  &=& \frac{\beta^2 + 2\theta_E^2}{\beta\sqrt{\beta^2 + 4\theta_E^2}}.
\label{mag1}
\end{eqnarray}
In this equation the angular distance of source with respect to the lens can be taken as a function of time. Assuming a constant relative velocity between the lens and source, without imposing 
any acceleration, the angular location of source in terms of impact parameter can be written as the combination of minimum impact parameter and relative angular velocity, $\mu$ as follows 
\begin{equation}
\beta^2 = \beta_0^2  + \mu^2 (t -t_0)^2,
\label{impact}
\end{equation}
where $\mu$ is taken a constant value for a simple lens. We can 
write equation (\ref{mag1}) in a simple form by dividing this equation to the Einstein angle, $\theta_E$ as follows: 
\begin{equation} 
A(t) = \frac{u^2 + 2}{u\sqrt{u^2 + 4}}, 
\label{pacf}
\end{equation}
where $ u = \beta/\theta_E$ and equation (\ref{impact}) changes to the following form of $u^2(t) = u_0^2 + (t-t_0)^2/t_E^2$. Here $t_E$ is the Einstein crossing time and is defined as $ t_E = \theta_E/\mu$. That is the typical variation of a microlensing light curve where the source star crosses the size of Einstein ring.

\begin{figure}[h]
\centerline{\psfig{file=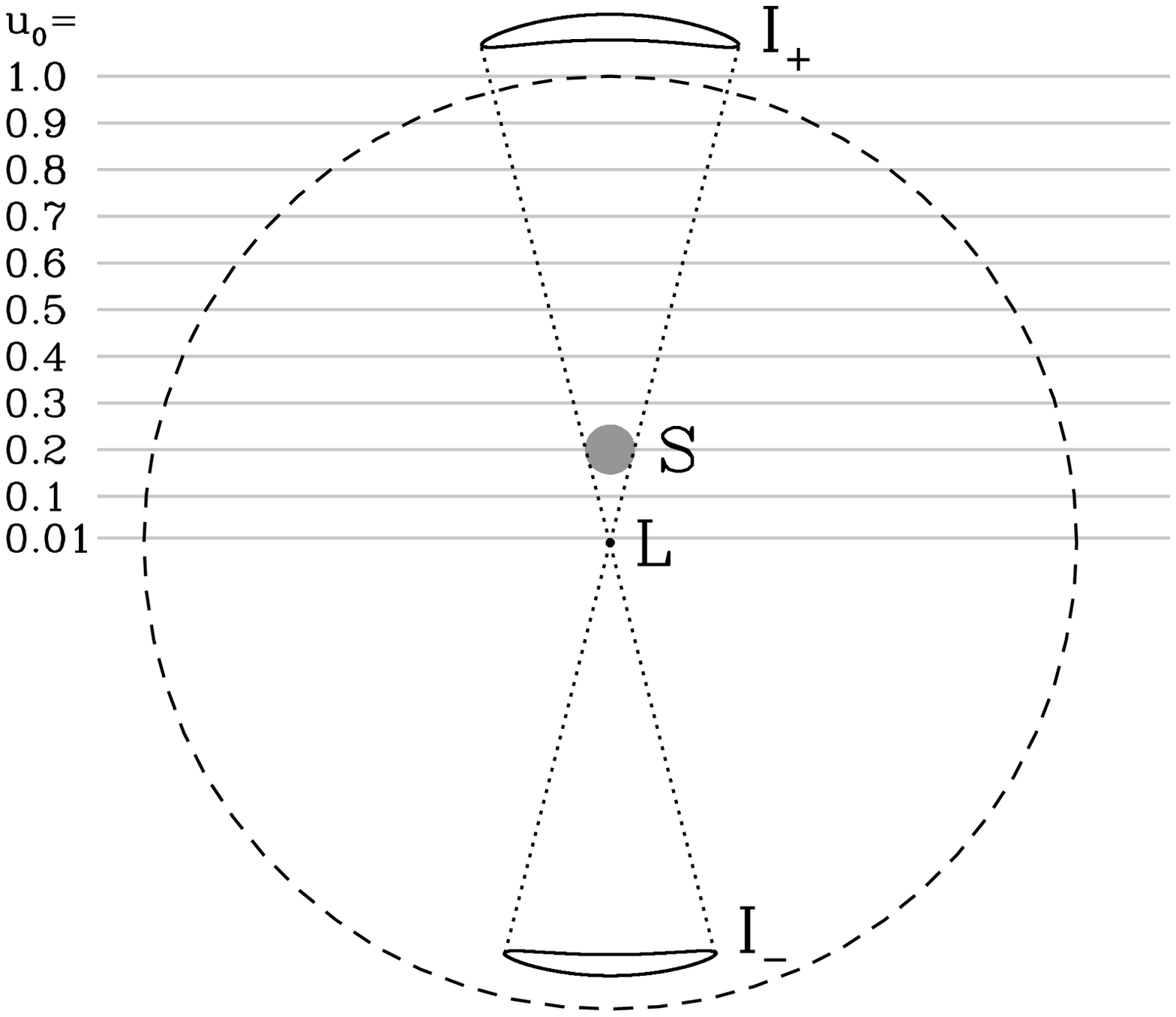,width=8.0cm}\psfig{file=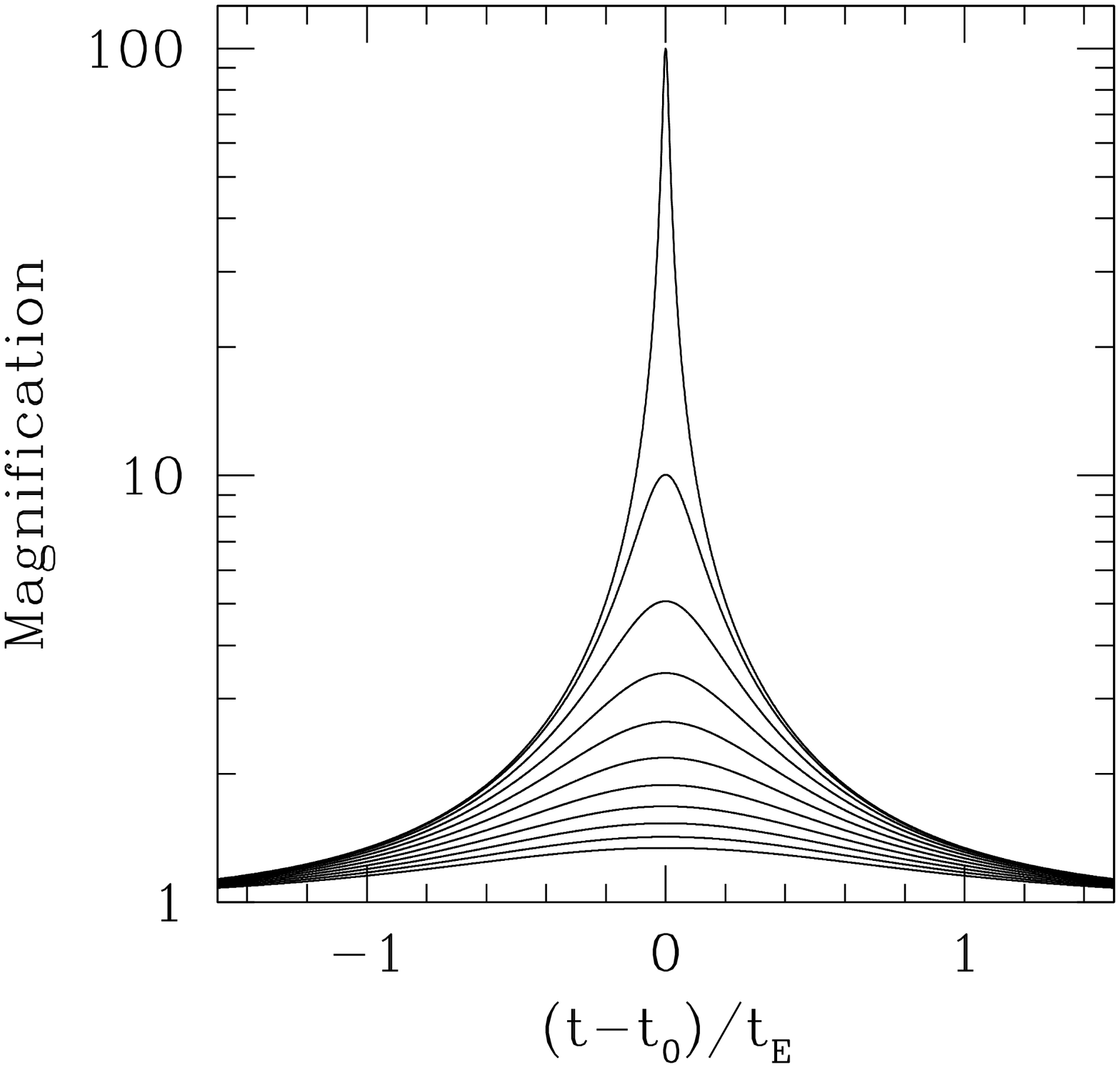,width=7.0cm}}
\vspace*{1pt}
\caption{Left panel represents the shape of images on the lens plane with point like lenses and different impact parameters. The right panel shows the magnification of source stars with different impact parameters which is obtained by the overall area of images divided to the area of source without magnification. This figure is adapted from \cite{gaudi,pac1996}}
\label{lightcurve1}
\end{figure}

Figure (\ref{lightcurve1}) shows gravitational microlensing light curve with different impact parameters. Smaller impact parameter results in higher magnifications of the light curve. 

The other observable parameter is the angular separation between the images and from equation (\ref{sol1}) this quantity is 
\begin{equation}
|\theta^+ - \theta^-| = \sqrt{\beta^2 + 4\theta_E^2}.
\label{sep}
\end{equation}
For the case that source, lens and observer are aligned on a straight line, the impact parameter is set to zero and angular distance between the images would be in the order of $2\theta_E$.  Another useful equation is writing the Einstein angle in simple form of 
\begin{equation}
\theta_E = \sqrt{\frac{2 R_{sch}}{D_s}\frac{1-x}{x}},
\label{aangle}
\end{equation}
where $R_{sch}$ is the Schwarzschild radius of lens, $x = D_d/D_s$. We note that the Einstein angle is a function of lens mass, distance of source and distance of lens from the observer. Using the numerical value of Schwarzschild radius, $R_{sch} = 2.95 (M/M_\odot)$ km, equation 
(\ref{aangle}) can be written as follows
\begin{equation}
\theta_E = 0.901 mas~
 (\frac{M}{M_\odot})^{1/2}(\frac{D_s}{10kpc})^{-1/2}(\frac{1-x}{x})^{1/2}. 
 \label{angle}
 \end{equation}
For the case that lens is located at the middle of distance between the observer and source (i.e. $x=1/2$) and for a typical lens with $ M_L = 0.5 ~M_\odot$, the Einstein angle for the the background stars in the Galactic bulge or Large Magellanic Clouds is about 
\begin{eqnarray}
\theta_{E,{\it Bulge}} & \simeq & 0.691~ mas, \\
\theta_{E,{\it LMC}} & \simeq &  0.285~ mas,
\end{eqnarray} 
which is smaller compare to the resolution of ground based telescopes. 

\begin{figure}[t]
\centerline{\psfig{file=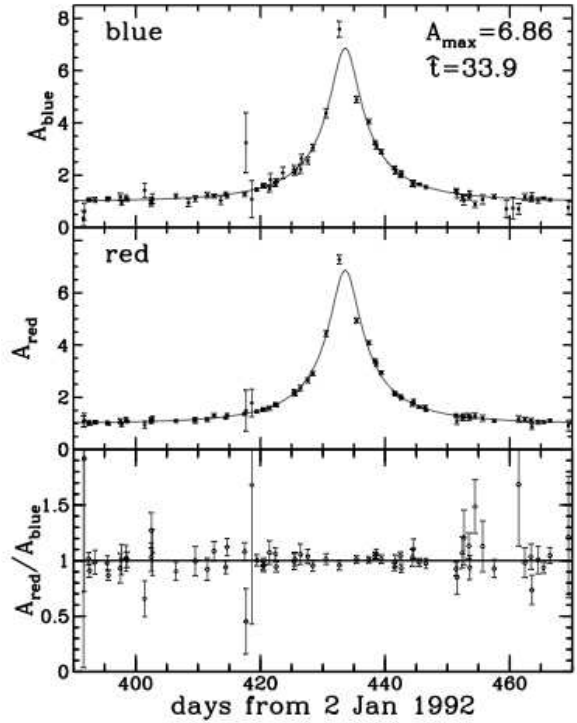,width=8.0cm}}
\vspace*{1pt}
\caption{The first gravitational microlensing candidate in the direction of Large Magellanic Cloud. The light 
curve has been observer in two blue and red filters. The lowest panel represents the ratio of best fit to the
microlensing light curve. Since gravitational microlensing is monochromatic phenomenon, magnification is 
independent of frequency of light \cite{macho1}.}
\label{fm}
\end{figure}

 For calculating the relative angular velocity of source with respect to lens, regarding that all three points of observer, lens and source are moving with respect to each other, we choose the position of observer as the centre of coordinate system. 
The relative velocities of the lens-Earth and source-Earth are given by ${\bf v_{LE}} = \bf{v_L} - \bf{v_E}$ and $\bf{v_{SE}} = \bf{v_S} - \bf{v_E}$, respectively. We set $\hat{\bf n}$ being a unit vector connecting Earth to the source star. A unit vector perpendicular to the line of sight can be made by combining velocity and vector $\hat{\bf n}$ as $$\hat{\bf l} = \frac{{\bf v}\times\hat{\bf n}}{|{\bf v}\times\hat{\bf n}|}.$$ The other unit vector orthonormal to $\hat{\bf n}$ and $\hat{\bf l}$ is $\hat{\bf m} = \hat{\bf n}\times\hat{\bf l}$. Substituting $\hat{\bf l}$, the unit vector $\hat{\bf m}$ is written as 
\begin{equation}
\hat{\bf m} = \frac{ {\bf v} - ({\bf v}\cdot\hat{\bf n}) \hat{\bf n}}{|{\bf v}\times\hat{\bf n}|}.
\end{equation}
Finally the projection of velocity on plane perpendicular to the line of sight and in the direction of $\hat{m}$ is 
given by $v_m = {\bf v}\cdot\hat{\bf m}$ or 
\begin{equation}
v_m = \frac{v^2 - (\hat{\bf n}\cdot {\bf v})^2}{|{\bf v}\times\hat{\bf n}|}.
\end{equation}
The relative angular velocity perpendicular to the line of sight is given by ${\bf \mu} = {\bf \mu_S} - {\bf \mu_L} $  as follows  
\begin{equation}
{\bf \mu} =  \frac{v_{SE}^2 - (\hat{\bf n}\cdot {\bf v_{SE}})^2}{D_s|{\bf v_{SE}}\times\hat{\bf n}|^2}\left( {\bf v_{SE}} - ({\bf v_{SE}}\cdot\hat{\bf n}) \hat{\bf n}\right) - \frac{v_{LE}^2 - (\hat{\bf n}\cdot {\bf v_{LE}})^2}{D_l|{\bf v_{LE}}\times\hat{\bf n}|^2}\left( {\bf v_{LE}} - ({\bf v_{LE}}\cdot\hat{\bf n}) \hat{\bf n}\right).
\end{equation}
For the perpendicular component of velocities along the line of sight (i.e. $v_\bot$),  
the relative angular velocity simplifies to 
\begin{equation}
{\bf\mu} = \frac{{\bf v_{S,\bot}} - {\bf v_{E,\bot}}}{D_s} - \frac{{\bf v_{L,\bot}} - {\bf v_{E,\bot}}}{D_l}.
\label{relative}
\end{equation}
Here, we assume that all three points are almost free particles at least within the time scale of lensing and velocities are constant vectors. In reality as the Earth is moving around the Sun, observer on the Earth is an accelerating point. We will consider this effect, so-called the parallax effect in the microlensing events. Using the typical velocity and distance for the source stars in the Galactic Bulge, the angular velocity is given by 
\begin{equation}
{\bf \mu} = 15.0 \mu as/day \left(\frac{D_s}{8.5 kpc}\right)^{-1} \left({\bf v_{S,\bot}} - {\bf v_{E,\bot}}- \frac{1}{x}({\bf v_{L,\bot}} - {\bf v_{E,\bot}})\right)\frac{1}{220 km/s}.
\end{equation}
This velocity depends on the line of sight and vary from the Spiral arms direction to the Galactic Bulge direction. Substituting this equation in the definition of Einstein crossing time, the time scale of gravitational microlensing events is
\begin{eqnarray}
t_E &=& 45.6 day \left(\frac{D_s}{8.5 kpc}\right)^{1/2}\left(\frac{M}{0.5 M_\odot}\right)^{1/2}\left(\frac{1-x}{x}\right)^{1/2}\nonumber \\
&\times& \left(|{\bf v_{S,\bot}} - {\bf v_{E,\bot}}- \frac{1}{x}({\bf v_{L,\bot}} - {\bf v_{E,\bot}}|\frac{1}{220 km/s}\right)^{-1}.
\label{te}
\end{eqnarray}
In the direction of Galactic bulge the typical time scale of $t_E$ is around one month. The first microlensing 
event has been reported in the direction of Large Magellanic Clouds by MACHO and EROS collaborations in 1993 \cite {macho1,eros1}. Figure (\ref{fm}) shows the light curve of this event in the red and blue filters. The observation 
was done in two different filters to distinguish the microlensing events from the variable stars, noting that gravitational 
lensing is independent of frequency of light and all the light rays follow the same geodesics in the space-time. 

From fitting 
the observational data of microlensing light curve with the theoretical light curve in equation (\ref{pacf}), we can 
obtain the three parameters of $u_0$ (minimum impact parameter), $t_0$ ( the time of maximum magnification) 
and $t_E$ (the Einstein crossing time). The only parameter that has physical information is $t_E$ and from equation (\ref{te}), this parameter is a function of mass of lens, distance of source star and lens from the observer and relative 
transverse velocity of lens and source with respect to the observer. In the microlensing experiments the main aim is to identify the physical parameters of lenses, however the degeneracy between the lens parameters makes this goal difficult. However, with the 
statistical methods for an ensemble of microlensing data, we can provide the distribution function for the lens parameters. The other methods for breaking degeneracy between the lens parameters is with including extra information as the motion of Earth around the Sun and finite-size effect of the source stars. We will discus extensively the second order effects later. 

\section{Statistics of Microlensing events}
\label{statistics}
The gravitational microlensing studies that initiated by Bohdan Paczynski in 1986 was for detecting massive compact halo objects of Galactic halo. The idea was monitoring huge number of stars in the Large and Small Magenllanic Clouds and looking for detection of microlensing events. Finally with counting the microlensing 
events we can estimate the contribution of lenses that compose the mass of Galactic halo.
\begin{figure}[h]
\centerline{\psfig{file=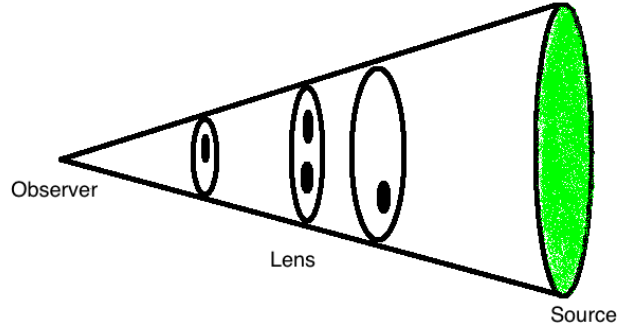,width=9.0cm}}
\vspace*{1pt}
\caption{Observer cone starting from the observer to the source stars at the background (painted by the green colour). Lenses along this cone cover the line of sight by the Einstein disks depicted by black circles. The ratio of overall angular area covered by the Einstein rings over the overall solid angle, represented by $\Omega$, provides the probability of microlensing event at 
a given instant of time.}
\label{cone}
\end{figure}

In order to estimate the probability of microlensing events in this direction, we can ask what fraction of stars in the Magellanic Clouds undergoes lensing if we take a snapshot from the stars in this field . Assuming that a microlensing event is effectively happen when a source crosses the Einstein ring of a lens, we can assign a cross section for the lenses with the size of $\pi R_E^2$ similar to the cross section calculations in the particle physics, where the Einstein radius is defined by $R_E = D_l \theta_E$. The observation is performed through a cone to the 
target stars at LMC as shown in Figure (\ref{cone}). Assuming density of lenses as a function of distance from the observer given by $\rho(z)$, the number of lenses within a differential element of $dz$ is given by $$ dN(z) = \frac{\rho(z)}{M} z^2\Omega dz,$$ where 
$\Omega$ is a solid angle towards target stars and $M$ is the typical mass of lenses. Multiplying $dN(z)$ to the 
area covered by the Einstein ring of lenses at the distance of $z$, represents the area of lenses that covers the area of our line of sight
towards the target stars, 
\begin{equation}
dA(z) =  \pi R_E^2 \frac{\rho(z)}{M} z^2\Omega dz.
\label{aa}
\end{equation}

By dividing $dA(z)$ to the area of 
cone at the position of $z$ (i.e. $S(z) = \Omega z^2$), the probability of 
lensing is given by 
\begin{equation}
d\tau(z) =  \pi R_E^2 \frac{\rho(z)}{M} dz.
\label{pp}
\end{equation}
We substitute the definition of Einstein radius 
$$R_E^2 = \frac{4GM_l D_s}{c^2}x(1-x),$$ in equation (\ref{pp}) 
and integrating along the line of sight, the overall optical depth is  
\begin{equation}
\tau = \frac{4\pi G D_s^2}{c^2}\int_{0}^{1}\rho(x) x (1-x) dx.
\label{tau}
\end{equation}
This probability similar to the propagation of particles in a given media in the particle physics is called the optical depth. An interesting point about this equation is 
that since area covered by the Einstein ring is proportional to the mass of lens, that cancels with the 
denominator in equation (\ref{pp}) and the optical depth is independent of the mass of lenses and 
it depends only on the density of lenses in the halo. 

In order to have an estimation from the optical depth, we calculate the contribution from the spherical 
structure of Galactic halo, using the density of halo as follows $$\rho(x) = \frac{dM}{4\pi D_s^3 x^2 dx}. $$ 
Substituting this equation in the definition of optical depth in (\ref{tau}), the optical depth is obtained in the 
order of  $\tau \simeq (v/c)^2$.  where $v^2 = GM/D_s$ is the rotation curve 
of Milky Way driven by the Galactic halo and is in the order of $220$ km/s. Here we assume that 
hundred percent of halo is made of MACHOs. Substituting the numerical value for $v$, the optical depth is 
$\tau \simeq 5\times 10^{-7}$ where the target stars are in the Magellanic Clouds. This means that observing $10^7$ stars in this direction, almost $5$ stars undergoes microlensing magnification. We will see later that from 
results of microlensing experiments, MACHOs just making 
less than $20$ percent of halo mass, then we expect to see only one event out of ten 
million stars in the Magellanic Clouds.

The other target for gravitational microlensing observations is the Galactic Bulge, where this direction has the advantage of large number of bright source stars and large column density of stars in the Spiral arms and Galactic Bulge. This direction has been used for exploring extra-solar planets orbiting around the lens stars. Also one 
of the important plans in this direction is using 
microlensing as the astrophysical tool for studying the physical properties of the source stars.  For this direction 
we can estimate the optical depth similar to the direction of Galactic halo. Here stars in the disk of 
Galaxy play both the role of lens and source stars. Assuming that disk has cylindrical symmetry, the mass of disk as a function of 
distance from the centre is given as $\rho(r) = dM/(2\pi h r dr)$ where $h$ is the width of the disk and $r$ is 
the distance from the centre of Galaxy. Substituting this density in equation (\ref{tau}), the optical depth obtain about 
$$\tau \simeq \frac{2 G M_{disk}}{c^2 h}.$$ For the case of stellar mass with 
$M_{disk} \sim 10^{10} M_\odot$, $h = 300$ pc, we get the value of $\tau\sim 3\times 10^{-6}$. A realistic value 
for the optical depth needs detailed calculations in different directions of Milky Way galaxy \cite{optd1,optd2}. 

The other statistical parameter for counting the microlensing events is the rate of events per a given time (traditionally 
given in terms of year). Let us assume that lenses are moving with constant transverse velocities and we monitor
a patch of sky for the duration of $T_{obs}$. In this case, each lens spans a rectangular area at the position of $z$ form the observer. The total area spans by lenses is 
\begin{equation}
d A(z) = 2 R_E v T_{obs}\times \frac{\rho(z)}{M_l} \Omega z^2 dz.
\label{rate1}
\end{equation}
We multiply the denominator and numerator of equation (\ref{rate1}) to $R_E$, moreover multiplying this equation
to the number of background stars. The result is the number of microlensing events for a given observation time of 
$T_{obs}$. Then, the rate of events per number of background stars and observation time is given by 
\begin{equation}
d\Gamma(z) = \frac{2}{\pi} \left({\pi R_E^2}\frac{\rho(z)}{M_l} dz \right)\times\frac{v}{R_E}.
\end{equation}
Comparing this equation with the definition of the optical depth in equation (\ref{pp}), the rate of events
can be written in the simple form of 
\begin{equation}
\Gamma = \frac{2}{\pi}\int \frac{1}{t_E}d\tau,
\end{equation}
or 
\begin{equation}
\Gamma = \frac{2}{\pi}\tau <\frac{1}{t_E}>.
\label{gammatau}
\end{equation}
This equation provides the rate of microlensing events in an  
ideal condition where all the ongoing microlensing events are observable for a telescope. In practice 
we miss very short duration microlensing events as well as events longer than the duration of 
experiment. The efficiency of detection of an event with the duration of $t_E$ is given 
by $\epsilon(t_E)$ function and provides fraction of events that can be detected by a telescope. This efficiency function obtain by synthetic observation, using a Monte-Carlo simulation and producing catalog from the microlensing data in 
the computer. Now, let us assume an experiment that detects $N$ microlensing events. Each event has an  
Einstein crossing time of $t_E^{(i)}$ where $i$ represents the index of event. 
The corrected rate of events in terms of efficiency function is given by the inverse of this function and divided to the exposure time and Number of background stars as follows
\begin{equation}
\Gamma = \frac{1}{N_{bg} T_{exp}} \sum_i^N\frac{1}{\epsilon(t_E^{(i)})}.
\end{equation}

The other important parameter that can be compared directly with the theory is the optical depth. We can 
express the geometrical definition of optical depth in terms of temporal definition, by means that optical depth is 
equal to the overall duration of microlensing events divided to the exposure--time times the number of background stars. The overall Einstein 
crossing time has to be corrected by the efficiency function, by means of $ \sum_i^N {t_E^{(i)}}/{\epsilon(t_E^{(i)})}$. Since the duration of events depends on the impact parameter, for the case of zero impact parameter the duration is $2t_E$ while 
for the impact parameter with $u_0=1$, according to the definition of an event with $A_{max}>1.34$, the duration is zero. For the impact parameter ranging form zero to one, the average impact parameter is $\pi/2\times t_E$. Then, the temporal optical 
depth is defined as 
\begin{equation}
\tau = \frac{\pi}{2}\frac{1}{N_{bg} T_{obs}}\sum_i^N \frac{t_E^{(i)}}{\epsilon(t_E^{(i)})}.
\label{tauexp}
\end{equation}
In what follows, we will compare the observed optical depth with the theoretical value.

\subsection{Observational results from detection of MACHOs}
The microlensing experiments of EROS\cite{e1,e2}, MACHO\cite{m1,m2,m3}, MOA\cite{mo} and OGLE\cite{o} have started observation of Large and Small Magellanic Clouds in 1990s for searching microlensing signals from compact objects as candidate for dark matter in the halo of Milky Way galaxy.  These telescopes were observing in survey mode by monitoring a large area of target stars and investigating the light curves for microlensing events. There were also small size follow-up telescopes for making full coverage of light curves with high cadence and higher signal to noise photometry data points. This mode of observation is still in used for extrasolar planet observations with microlensing method.
 
After producing photometric images, the main part of observation is producing light curve from the images. Especially in the crowded fields it is difficult to distinguish the time variation of background stars from the microlensing events. The observation groups developed 
image difference method where change  in the flux of a star in the crowded fields can be identified by differencing images from a template image of the field\cite{diff}. The result is identification of time variation of a star as a function 
of time. Due to variability of source stars, there are large number of backgrounds that has to be cleaned from the 
 signals. In order to select microlensing signals, stable stars are chosen for flux variation analyzing and in the direction of Magellenic Clouds, out of millions of background stars few microlensing candidate 
 has been found by the observation groups.

 The next step is analyzing light curves by fitting simple microlensing 
 light curve and extracting the parameters of events. The only physical parameter in the light curve of single lens 
 is the Einstein crossing time which depends on three parameters of lens mass, distance of lens and source from the observer and transverse velocity of lens. Due to degeneracy in the parameters, we need to perform statistical analysis in terms of number and duration of events.  Since the number of events in the direction of LMC/SMC is small, there would be large Poisson error. The final number that will help us to estimate the contribution of lenses in the Galactic halo is from the optical depth calculation. An extensive discussion on the optical depth calculation from 
 the experiment can be found in Ref \cite{optd2}. The most challenging problem in the optical depth calculation is the 
 blending effect. This effect results from the mixture of background stars within 
 the Point Spread Function (PSF) of source star.

The blending effect can have two important effects of (i) modifying the shape of light curve that provides a wrong estimation for the value of Einstein crossing time of a microlensing event and (ii) we get wrong estimation for the 
number of background stars in the field. In the blending effect out of many stars inside PSF one of the stars in PSF disk is magnified, then the overall flux is given by 
\begin{equation}
F(t) = F_0 A(t) + F', 
\end{equation}
where $F_0$ is the microlensed star and $F'$ is the overall flux of background stars. Then the observed magnification 
is given by 
\begin{equation}
A_{obs}(t) = \frac{ F_0 A(t) + F'}{ F_0 + F'}.
\end{equation}
We can rewrite this equation in the following form of 
\begin{equation}
A_{obs}(t) = b A(t) + (1-b),
\end{equation}
where $b$ is the blending parameter and is given by $b = F_0/(F'+F_0)$ and ranges between zero for when 
there is maximum blending and one when there is no blending. For the case of $b<1$, the observed magnification 
is always smaller than intrinsic magnification (i.e. $A_{obs}<A$). This means that for a given impact parameter of $u_0$, the blended microlensing event will suppressed and the result would be smaller Einstein crossing time 
for a microlensing event. One the other hand since $t_E\propto \sqrt{M}$, then estimated mass for the lens
would be smaller than real value. 

\begin{figure}[h]
\centerline{\psfig{file=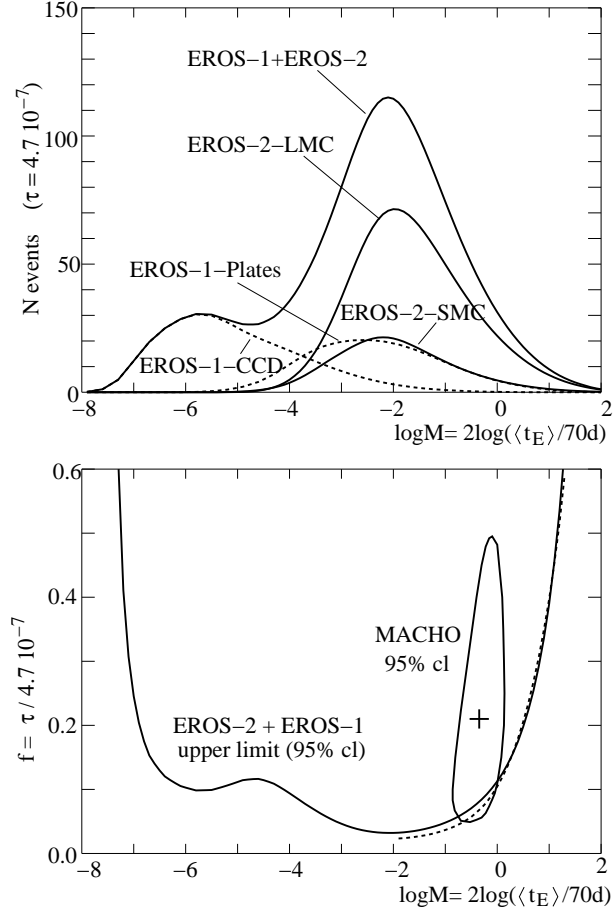,width=8.0cm}}
\vspace*{1pt}
\caption{For the Standard model of Galactic halo \cite{mtau}, the expected distribution 
of events in terms of mass of lenses in the upper panel and excluding halo made of MACHOs for a given lens mass is at the lower panel. This Figure is adapted from Ref. \cite{etau2}.  }
\label{o1}
\end{figure}
The other problem of blending is that we will get an incorrect number for the source stars in the background. Since the optical depth from the experiment in equation (\ref{tauexp}) is given in terms of the number of source stars, underestimating background stars will provide overestimating the numerical value for the optical depth. The microlensing 
experiments updated their measured optical depth by discarding some of their microlensing candidates as well as 
having better value for the number of background stars from blending estimation. The updated value for the optical depth of MACHO group in the direction of LMC\cite{b1,b2} is $\tau = (0.4 \pm 0.1) \times 10^{-7}$  after 5.7 years of collecting data 
\cite{mtau} while the first estimation was a larger number of $\tau = 1.2^{+0.4}_{-0.3} \times 10^{-7}$ . The EROS collaboration also from $6.7$ years data collection \cite{etau1} found three microlensing candidate towards LMC and 
obtained the optical depth of  $\tau = (0.15\pm0.12)\times 10^{-7}$. In order to decrease the blending effect, EROS 
group also used bright stars for estimating the optical depth. In the direction of LMC, for the bright sources there 
was one microlensing candidate and using this subsample they obtained the following limit of $\tau<0.36\times 10^{-7}$
within $95\%$ of level of confidence \cite{etau2}. Figure (\ref{o1}) represents an upper limit to the contribution of MACHOs in the Galactic halo. Using the Standard model of Galactic halo \cite{mtau}, the expected distribution 
of events in terms of mass of lenses is in the upper panel and excluding halo made of MACHOs for a given lens mass is at the lower panel of Figure (\ref{o1}) \cite{etau2}. This analysis shows that within the range of $10^{-6}$ to few solar mass MACHOs are making less than 
$20\%$ of halo mass.  In this analysis the mass function of lenses are taken as Dirac-Delta mass function, given in the X-axis of Figure (\ref{o1}).

The other method for estimating blending parameter is using images from the Hubble Space Telescope ({\it HST}) to compare some parts of sub-fields of target stars with 
diffraction limit images \cite{mr,beros,han}. The other possibility is 
using Lucky imaging method to remove the turbulent effects of atmosphere and producing least blended images 
from the field of stars \cite{L1,L2}.

\section{Second order effects in single microlensing events}
\label{perturbation}

We have seen that for the case of microlensing events with the single lens, the Einstein crossing time which is the only physical 
parameter of lensing, depends on the mass, distance and transverse velocity of lens with respect to the line of sight.  Due to this degeneracy we can not identify the parameters of an individual event. However, as we will see in this section, the second order effects as parallax and finite size of source star will help us to break partially the degeneracy between the lens parameters.

\subsection{ Parallax Effect}
\begin{figure}
\centerline{\psfig{file=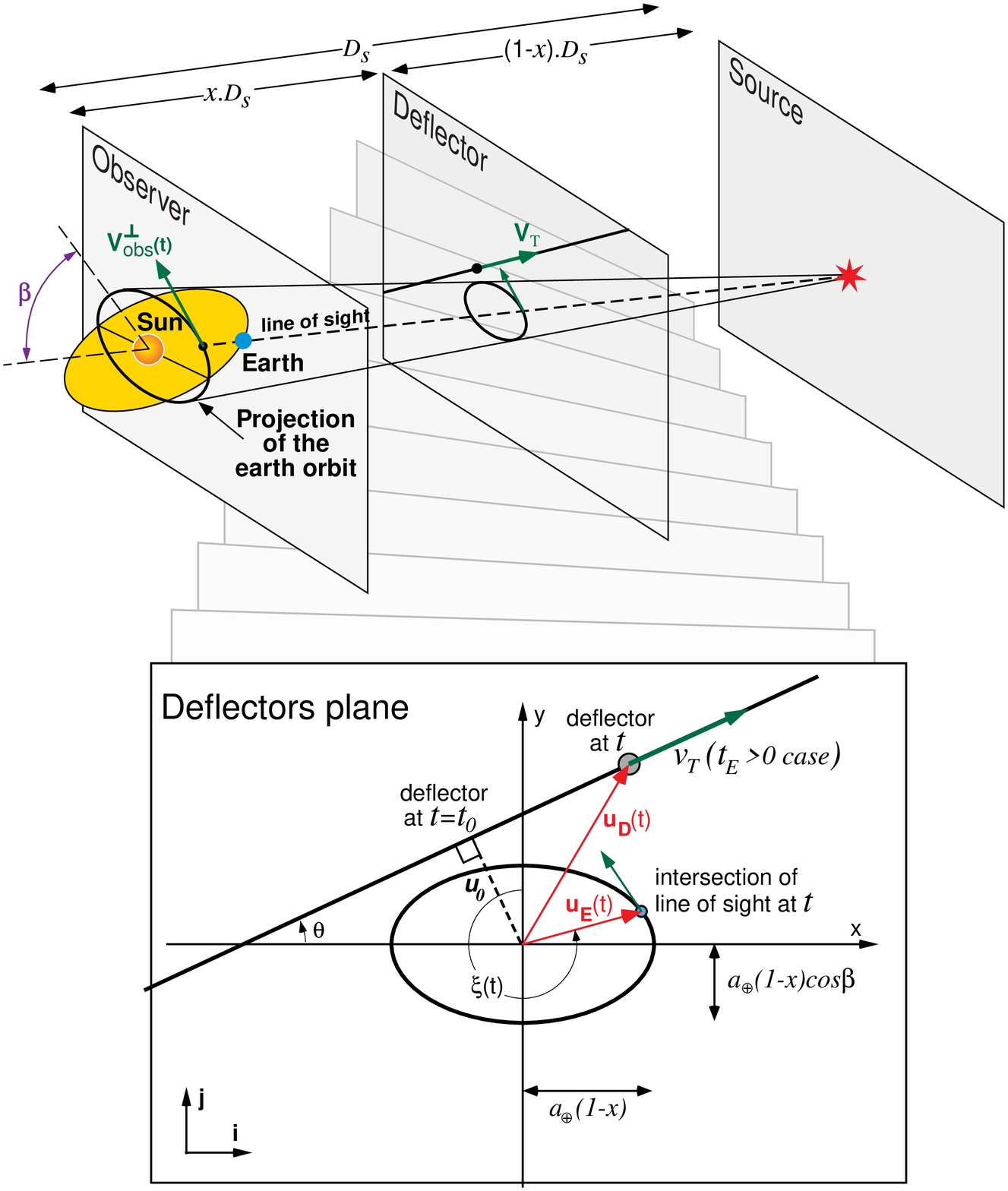,width=8.0cm}}
\vspace*{1pt}
\caption{Relative motion of observer with respect to the lens on the lens plane. In the upper panel the line of 
sight to the source star crosses the lens plane while Earth is moving around the Sun and 
this intercepted line follows an ellipse on the lens plane. On the other hand lens follows a straight line. Here 
the position of Sun is located at the centre of ellipse and relative distance between the line of sight to the source and 
lens enters in the single magnification formula of equation (\ref{pacf}). Figure is adapted from Ref \cite{rahvar2003}.}
\label{p1}
\end{figure}

In simple microlensing light curve, the relative motion of observer--lens--source in equation (\ref{relative}) is non-accelerating. Any accelerating motion of these three points results in deviation of light curve from simple microlensing light curve. The simple microlensing light curve is symmetric as shown in Figure (\ref{lightcurve1}). The annual motion of Earth around the Sun or rotation of lens around companion or source around companion 
results in deviation of light curve from single microlensing light curve. Figure (\ref{p1}) demonstrate the rotation of Earth around the Sun and subsequently the relative motion of observer-lens-source \cite{rahvar2003}.

In order to calculate the relative velocity, we first calculate the coordinate of all components of observer, lens and source, then derive velocity from the equation of motion. The motion of earth around sun, projected along the line of sight is give by
\begin{equation}
\vec{r}_E(t) = a_\oplus \cos\xi(t) \hat{i} + a_\oplus \cos\beta\sin\xi(t) \hat{j},
\label{earth}
\end{equation}
where Sun is located at the centre of coordinate and $a_\oplus$ is the orbital radius of Earth, $\xi(t)= \omega t + \phi_0$ is the orbital phase of Earth, $\omega$ is the orbital angular velocity of Earth and $\beta$ is the angle between the ecliptic and deflector plane. Here we assume negligible eccentricity for the earth orbit. Hence the velocity 
of earth from derivation of equation (\ref{earth}) is given by 
\begin{equation}
\vec{v}_E(t) = - \hat{i} a_\oplus \omega \sin\xi(t)  +  \hat{j} a_\oplus \omega \cos\beta\cos\xi(t). 
\end{equation}
The position of lens projected on the lens plane can be given as follows:
\begin{equation}
\vec{r}_L(t) = \hat{i}R_E\left[\frac{t-t_0}{t_E}\cos\theta_L + u_0\sin\theta \right]  + \hat{j}R_E\left[\frac{t-t_0}{t_E}\sin\theta_L +u_0\cos\theta \right],
\label{lens}
\end{equation}
where $\theta_L$ is the angle of trajectory of lens with respect to the semi-major axes of earth orbit, projected on the lens plane and $\theta_L\in[0,2\pi]$. We note that 
in order to cover all the possible configurations, the velocity of lens needs to be both positive and negative signs. Hence
we let $t_E$ has both positive and negative signs. The time 
derivation from equation (\ref{lens}) results in velocity of lens as follows
\begin{equation}
\vec{v}_L(t) = \hat{i}\frac{R_E}{t_E}\cos\theta_L +\hat{j}\frac{R_E}{t_E}\sin\theta_L,
\end{equation}
where $v_L= R_E/t_E$ is the transverse velocity of lens (and can has positive and negative signs). Similar argument can be used for the motion of 
source star.  The velocity of source is given by 
\begin{equation}
\vec{v}_S(t) = \hat{i}v_S \cos\theta_S +\hat{j}v_S\sin\theta_S,
\end{equation}
where $v_S$ is the projected velocity of source star parallel to the lens plane and $\theta_S$ is the angle with respect to the semi-major axis 
of Earth orbit. The apparent relative velocity of these three points is given by equation (\ref{relative}).  In more complicated case, in addition to the observer, the source star and lens can undergo an accelerating motion, if they belong to gravitationally many-body bounded system. The case for a 
binary lenses and binary sources can be found in Ref \cite{rah,matt}. 
For simplicity, we assume that the motion 
of lens and source are non-accelerating and contribution from the annual motion of earth introduces time variation
in the relative velocity of these three points.

In order to calculate the magnification of source star from equation (\ref{pacf}), we need the relative distance between the observer--source intercepting line in the lens plane with the position of lens. We use equations (\ref{earth}) and (\ref{lens}) 
to calculate this distance. In order to calculation magnification from equation (\ref{pacf}), all the scales should be projected on the lens plane and normalize to the Einstein radius. The relative distance on the lens plane is given by ${\bf  u(t)}$ vector as follows
\begin{eqnarray}
{\bf u(t)} &=&  {\bf \hat{i}} \left( \delta u \cos\xi(t) - \frac{t-t_0}{t_E}\cos\theta_L - u_0\sin\theta_L\right)\nonumber \\
& +& {\bf \hat{j}} \left( \delta u \cos\beta\sin\xi(t) - \frac{t-t_0}{t_E}\sin\theta_L - u_0\cos\theta_L\right),
\label{ut}
\end{eqnarray}
where $\delta u = {a_\oplus (1-x)}/{R_E}$. In literature, $\delta u$ is also expressed by the symbol of $\pi_E$ \cite{gould92} also the projected velocity of lens on the observer position, $\tilde{v} = v/(1-x)$. These two parameters are related to each other by
\begin{equation}
\tilde{v} =  \frac{a_\oplus}{ t_E \delta u}
\end{equation}

\begin{figure}[h]
\centerline{\epsfig{file=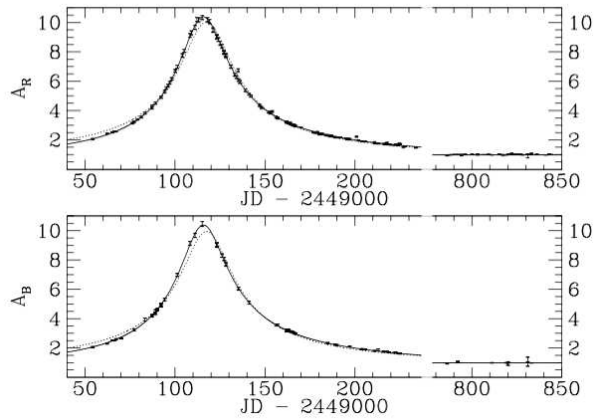,width=8.0cm}}
\vspace*{1pt}
\caption{First microlensing event with parallax signature. The effect is deviation of light curve from symmetric simple microlensing shape. The upper panel is magnification verses time in R-band and lower panel is at B-band. The parameters of lens are $t_E= 110$ days and projected velocity of lens on solar position $\tilde{v} = v_L/(1-x) = 75\pm 5$ km/s or $\delta u = 0.21$. Figure is adapted from 
Ref \cite{alpar}.}
\label{par2}
\end{figure}
Figure (\ref{par2}) shows the first parallax microlensing event observed by MACHO group \cite{alpar}. The parallax effect causes simple microlensing light curve deviates from its symmetric shape. For the case of Figure (\ref{par2}), the Einstein crossing time is 
$t_E \simeq 110$ days, comparable to one year annual motion of earth. Having longer events, the trajectory for the impact parameter, 
${\bf u(t)}$ follows more curved path compare to the short duration events. Hence for events with $t_E$ in the order of hundred days, the parallax effect is more effective. 

\begin{figure}[h]
\centerline{\psfig{file=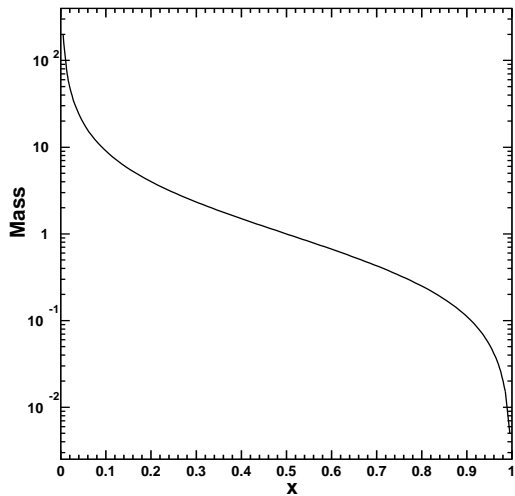,width=8.0cm}}
\vspace*{1pt}
\caption{Relation between the mass of lens (in solar mass unit) verses distance of lens from the observer in terms of 
$x = D_l/D_s$. The parameter of $\delta u = 0.2$ is assumed to be measured from the microlensing light 
curve.  Here we set source star is located at $D_s =8$ kpc in the Galactic bulge.}
\label{figdu}
\end{figure}
From equation (\ref{ut}), the extra parameters that enter in calculation of $u(t)$ are $\delta u $ and $\theta_L$ which 
can be obtained from fitting theoretical light curve (including the parallax effect) to the observed data. From the five parameters of $u_0$, $t_0$, $t_E$, $\theta_L$ and $\delta u $ enters to the magnification function, only $t_E$ and $\delta u$ have physical information. From simple lensing, we have seen that knowing $t_E$, we obtain a constrain between the mass, distance and relative velocity of the lens from the observer. The parallax parameter, $\delta u$ has an extra information and from equation (\ref{ut}), $\delta u$ can be written as
\begin{equation}
\delta u  = a_\oplus (\frac{4GM}{c^2}D_s)^{-1/2}(\frac{1-x}{x})^{1/2},
\end{equation}
unlike to the Einstein crossing time this parameter depends only on the mass 
and distance of lens from the observer. Figure (\ref{figdu}) shows constraining relation between the mass
of lens and relative distance of lens and source for a given value of $\delta u = 0.2$. In Figure (\ref{figdu}), we 
assume source is located at Galactic bulge, where $D_s\sim 8$ kpc. We can estimate the mass and distance of lens from the observer if we use models for the distribution of mass in the Milky Way galaxy and the mass function of lenses \cite{ps}.

\subsection{Simultaneous parallax measurement by a ground-based telescope and a telescope orbiting 
around the Sun}
Traditionally, the displacement of angular position of an object compare to the reference points at infinity by 
two different observers is also called parallax effect.  This method has been used in the astronomy for 
measuring the distance of remote stars by measuring the displacement of the angular position of a given star 
relative to the background stars at two different times, preferably with six month difference. We use this technique for observation of gravitational microlensing events, but since microlensing is a transient event, we need to perform observation at the same time while from two different positions.

\begin{figure}[h]
\centerline{\psfig{file=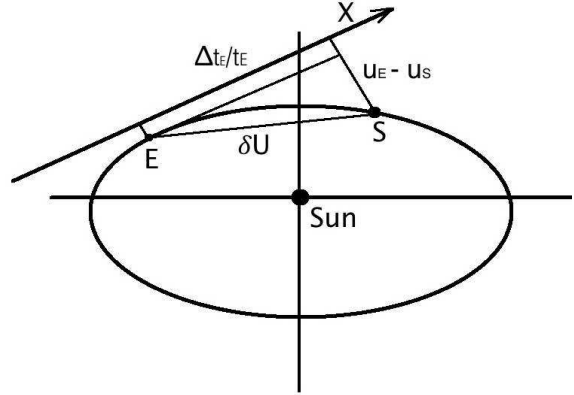,width=8.0cm}}
\vspace*{1pt}
\caption{Relative distance of Earth and a satellite orbiting around the sun from the position of lens. The ellipse 
is the projected trajectory of earth on the lens plane, $u_E$ and $u_S$ are the minimum distance of earth and satellite 
with respect to the lens, respectively. $\delta U$ is the distance of the Earth and satellite normalized to the Einstein radius at the maximum magnification for light curve of each observer. }
\label{p2}
\end{figure}

The degeneracy between the lens parameters in the microlensing observation has been known since 50 years and for the first time Refsdal and Liebes proposed breaking this degeneracy by observation of microlensing events from two different positions. One of the observers is located on the Earth and the other one is a satellite telescope, orbiting around the Sun \cite{refsdal1,refsdal2,Liebes}.  Let us assume that the projected distance between the Earth and satellite on a plane parallel to the lens plane, at the position of solar system is 
$d_\bot$. From the observation of light curve we can identify the minimum impact parameters as well as the time difference between the maximum peak of the light curves for the two observers. Lets us use $u_{E}$ for minimum impact parameter of the Earth observer, $u_{S}$ for minimum impact parameter for the satellite observer and $\Delta t$ for time difference between the maximum magnification for the two observers.  The projection of $d_\bot$ on the lens plane normalized to the Einstein radius 
is related to the difference in the impact parameters and time difference between the peaks as follows:

\begin{equation}
(\delta U)^2= (\frac{\Delta t}{t_E})^2 + (u_{E}-u_{S})^2,
\label{comp}
\end{equation} 
where $\delta U = (1-x) {d_\bot}/{R_E} $. The geometrical details are depicted in Figure (\ref{p2}). The relation 
between $\delta U$ and $\delta u$ or $\pi_E$ that is introduced in previous section is given by $\delta U =  \pi_E \times 
d_\bot/a_\oplus$. Hence the parallax parameter in terms of two components in equation (\ref{comp}) 
is given by 
\begin{equation}
{\bf \pi_E} = \frac{a_\oplus}{d_\bot}(\frac{\Delta t}{t_E},u_E - u_S), 
\label{pie}
\end{equation}
where ${\Delta t}/{t_E}$ is along x-axis, parallel to the trajectory of the lens 
and y-axis is perpendicular to it. 

From Figure (\ref{p2}), we note that since $u \simeq t/t_E$ (i.e. impact parameter is proportional to time over the typical lensing time), then $|u_E - u_S| \simeq \Delta t/t_E$ and we would expect that two components of $\pi_E$ vector to be in the same 
order \cite{Novati}. Also we can identify the trajectory of source star in the lens plane by $\tan\alpha = (u_E - u_S)t_E/\Delta t$. Recently it has been possible to observer simultaneously the microlensing events
from the Earth and Spitzer Satellite orbiting around the Sun \cite{Novati}.  Dedicated telescopes orbiting around the Sun 
in future will break partially the degeneracy of microlensing parameters. Using $t_E$ form the light curve of event and the average value for the relative transverse speed of lens, we can estimate the Einstein radius. Combining with equation (\ref{comp}), we can calculate the mass and distance of lens from the observer. The main impact of this kind of projects 
would be identifying the distance of lenses, their masses and more impotently the 
distance of planets orbiting around the lens stars.

\subsection{Terrestrial Parallax effect}
The other kind of microlensing parallax observation is the terrestrial parallax. The parallax observation from Earth 
has been used for long time for measuring distance of close-by objects such as moon. However, the accuracy of parallax observation from the earth for distant objects is very low and practically it is useless. In the case 
of very high magnification microlensing events, observers at different positions on the Earth can see the peak of light curve at different times. Using this argument that $\Delta u \simeq \Delta t/t_E$,  the time difference between the observation of peaks for two observer would be $\Delta t \simeq t_E \times \Delta u$. Taking $t_E \simeq 10$ days, $\delta u = D_\oplus/R_E$ where $D_\oplus\simeq 10^4$ km is the diameter of earth and $R_E\sim 1$ A.U., we get $\Delta t \simeq 1$ min. The first terrestrial parallax microlensing event has been 
observed in OGLE-2007-BLG-224 event where the reported time delay in peak of microlensing light curve 
between three observatories of  South Africa,  Canaries and Chile were about one minute \cite{gpar}.

\subsection{Finite source effect}
The other important perturbative effect compare to the simple microlensing light curve is the finite-source effect.  The finite-source effect happens when the size of source projected on the lens plane is not negligible compare to the 
impact parameter of lens from the centre of source star. Then the different parts of source will magnify with different factor and result would be deviation of light curve from the simple lensing. 

\begin{figure}[h]
\centerline{\psfig{file=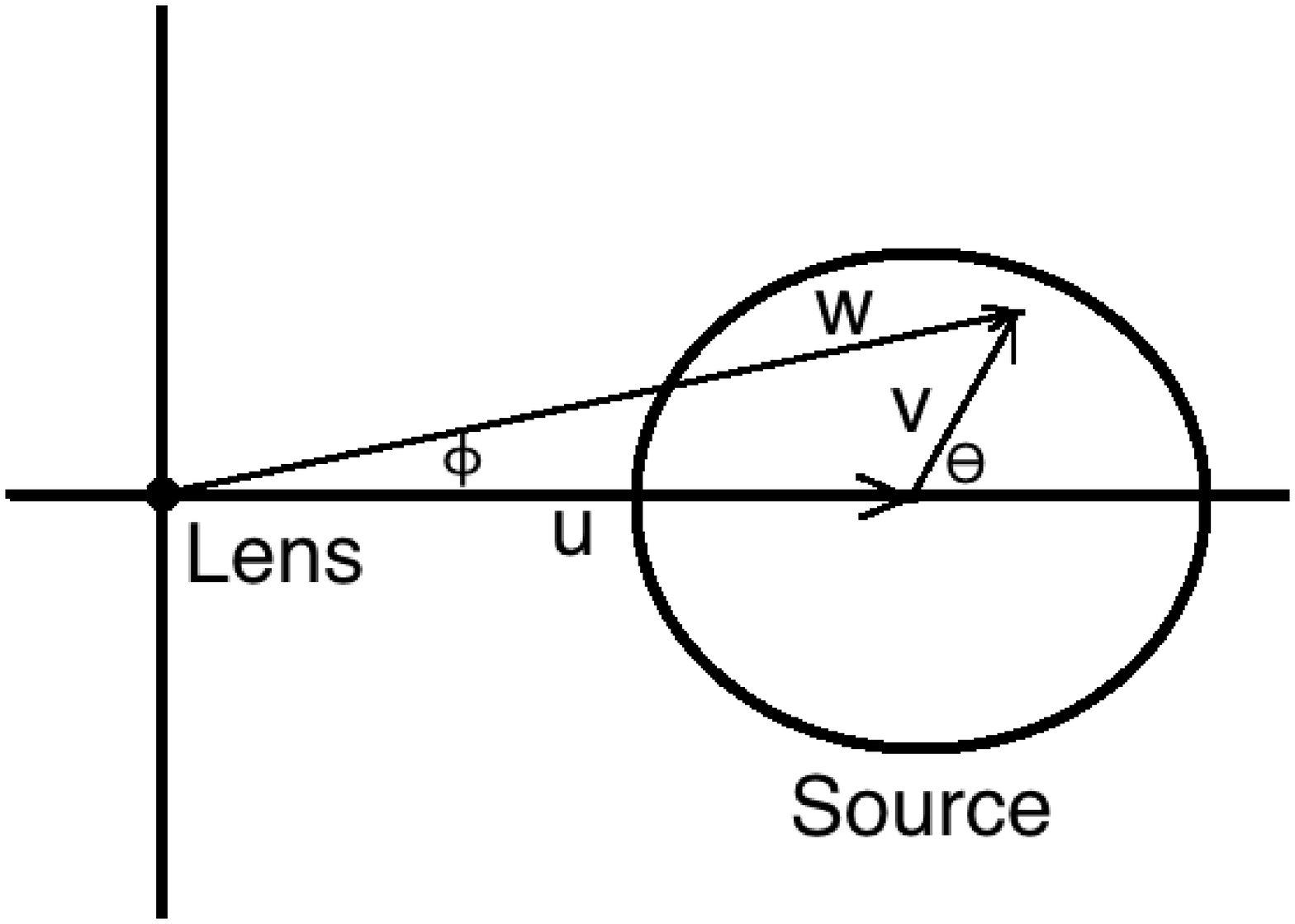,width=8.0cm}}
\centerline{\psfig{file=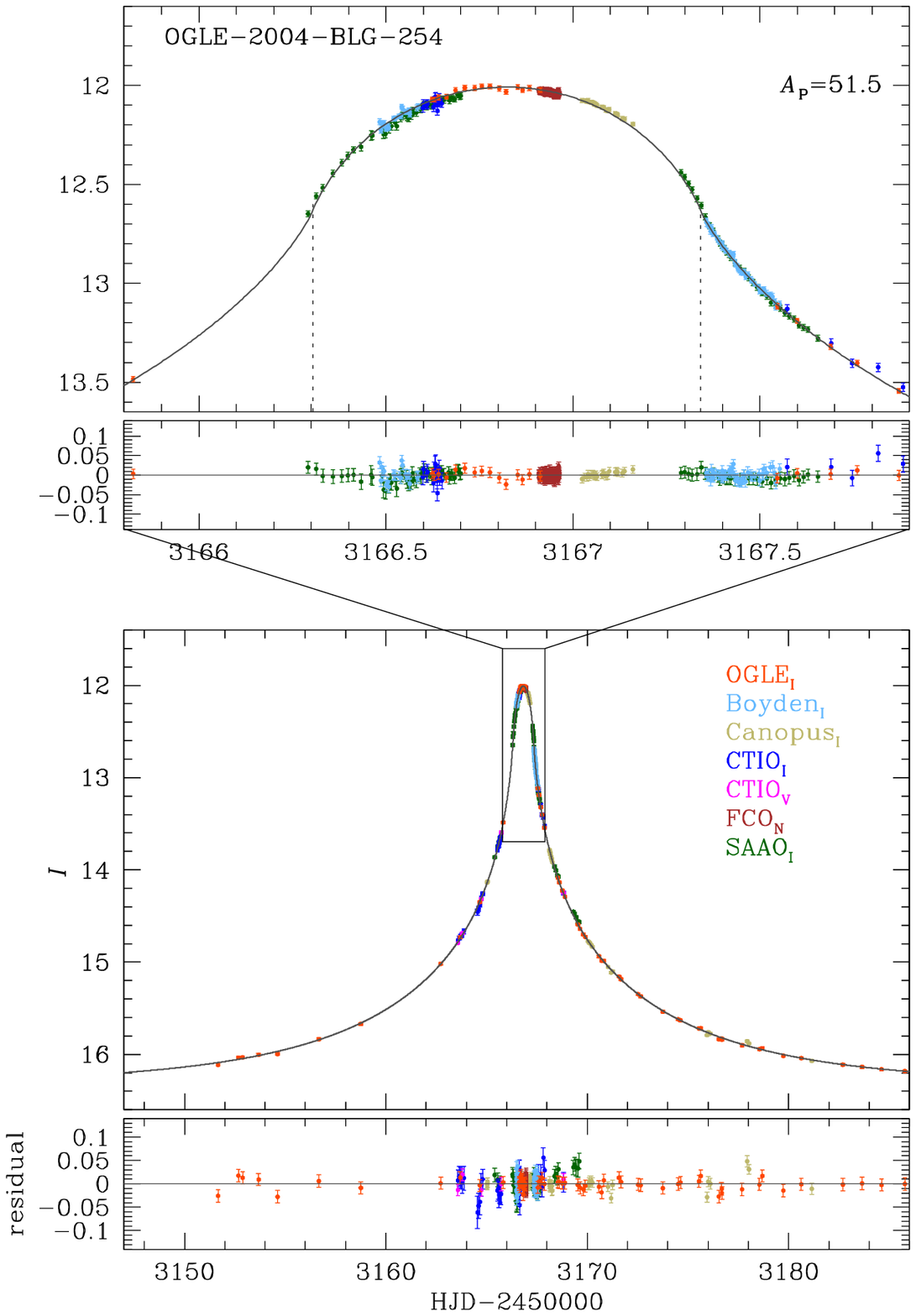,width=8.0cm}}
\vspace*{1pt}
\caption{Upper panel demonstrates the distance of projection of source star on the lens plane from lens. The lower
panel is a sample of very high magnification event OGLE-2004-BLG-254 and the finite source effect on the 
light curve \cite{chan}.}
\label{fsize}
\end{figure}

Let us assume that flux of star is also non-uniform while disk of star is rotationally symmetric around the centre of disk. We choose the coordinate at the centre of disk of source star, using the magnification factor depends on the 
distance of lens from each point of source disk. The configuration is shown in upper panel of 
Figure (\ref{fsize}) and the overall magnification factor is given by 
\begin{equation}
A(u) = \frac{\int_{\theta =0}^{2\pi}\int_{v =0}^{\rho_\star} A(w)F(v)v dv d\theta}{\int_{\theta =0}^{2\pi}\int_{v = 0}^{\rho_\star} F(v) v dv d\theta},
\label{feq}
\end{equation}
where $w = (u^2 + v^2 + 2uv\cos\theta)^{1/2}$ and $\rho_\star$ is the size of source star projected on the lens plane, normalized to the Einstein radius as shown in Figure (\ref{fsize}). Detailed analytical integration of equation (\ref{feq}) for a source with constant flux over its surface, is calculated by Witt and Mao\cite{wmao} in 1994. In this case we can simplify 
two dimensional integral to a one dimensional integral. Let us change coordinate system from $(u,\theta)$ to 
$(w,\phi)$ where $\phi$ is angle between $w$ and a line connecting lens to the centre of source star and is given by $$\phi = \arccos (\frac{u^2+w^2 - {\rho_\star}^2}{2uw}).$$
Then the differential element for area is given by $wdwd\phi$. In this case for three cases of $w>u$, $w<u$ and $w=u$, the magnification factor is given by 

\begin{eqnarray}
A(u) &=&  \frac{2}{\pi\rho_\star^2}\int_{|u-\rho_\star|}^{|u+\rho_\star|} \frac{2+w^2}{\sqrt{4+w^2}} \arccos (\frac{u^2+w^2 - {\rho_\star}^2}{2uw}) dw\nonumber \\
 &+& \frac{\pi}{2}H(\rho_\star - u) (\rho_\star - u)\sqrt{(\rho_\star - u)^2 +4 },
\end{eqnarray}
where $H$ is Heaviside function and is equal to one if $u<\rho_\star$, otherwise it is zero. The maximum magnification in finite-size effect at $u=0$ is given by $A_{max} = \sqrt{1+4/\rho_\star^2}$ which is larger for the small size stars. For the 
case of $\rho_\star\rightarrow 0$, we will have singularity in the maximum magnification.

Figure (\ref{fsize}) (lower panel) shows a sample of light curve with finite source effect for a very high 
magnification microlensing event \cite{chan}. One of the advantages of high magnification events is that lens 
can probe the luminosity of surface of source star. Since the luminosity due to limb darkening is not uniform and 
the apparent luminosity of stars decreases around the edge of star, the gravitational microlensing is a 
unique tool to probe the limb-darkening of stars few thousand parsec far from us.  A simple limb-darkening model is 
$I = I_0\left[1 - u_\lambda(1-\cos\phi)\right]$, where $\phi$ is the angle between the line of sight and a vector normal to the surface of the star. The best value of limb-darkening parameter in Figure  (\ref{fsize}) for event OGLE-2004-BLG-254  is $u_R = 0.70\pm 0.05$ and $u_I = 0.55\pm 0.05$ where the size of star is $\rho^\star = 0.0400\pm 0.0002$. The parameters of limb darkening can be measured for any high magnification event with a perfect coverage of the peak of light curve. One of these samples is OGLE 2008-BLG-290 \cite{pascal}

\subsection{Detection of stellar spots in Finite size effect}
\begin{figure}[h]
\centerline{\psfig{file=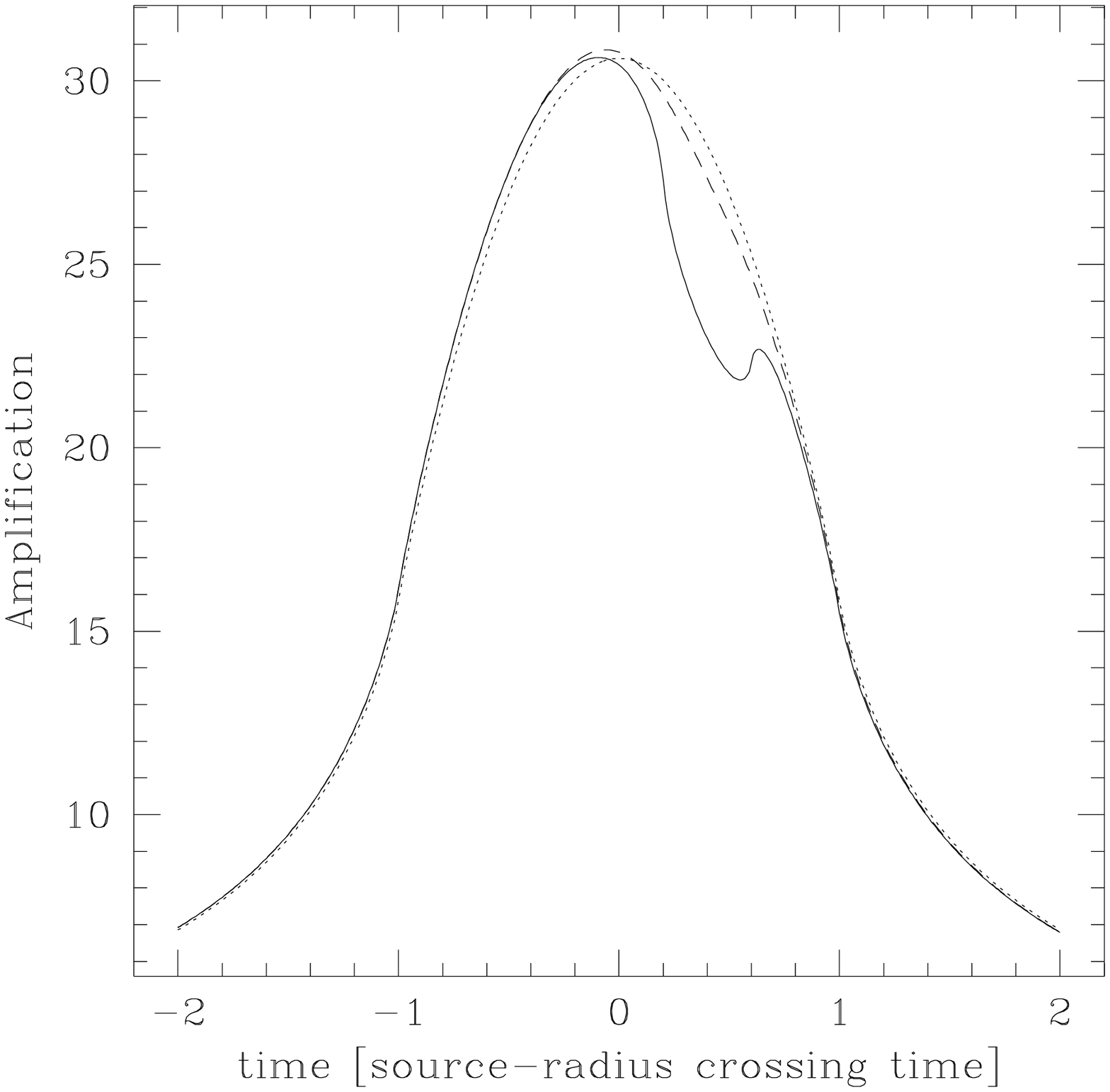,width=7.0cm}}
\vspace*{1pt}
\caption{Simulation of a microlensing light curve with zero impact parameter and a spot with the 
size $r_s = 0.2$, normalized to the radius of the star. The solid line and dashed line are with different distance of 
spot from the lens and dotted line is the microlensing light curve without spot. This figure 
is adapted from Ref \cite{dspot}. }
\label{fsample}
\end{figure}

The capability of gravitational microlensing as a natural telescope can also be used for probing the stellar spots on the surface of stars located at the other corner of Milky Way galaxy \cite{dspot,sr2}. The presence of spot produces perturbations in the microlensing light curve when lens get closer to the spot. We expect to observer such an effect in very high magnification events when the impact parameter is in the order of projected size of star on the lens plane.  In this 
case the magnification on the source star is given by 
\begin{equation}
A(r) = \frac{\int_S F(r') A(|r-r'|)d\Sigma' - \int_{S'} F_D(r') A(|r-r'|)d\Sigma'}{\int_S F(r') d\Sigma' - \int_{S'} F_D(r') d\Sigma'},
\end{equation}
where $r$ is the distance of centre of source from the lens and $r'$ is the position of points on the surface of source 
from the centre of source disk. $F(r)$ is the spotless flux of source and $F_D(r) = F(r) - F_S(r)$ is the flux 
increment between the spotless source and flux of spots.  The integral index of 
$S$ represents overall area of source and the index of $S'$ represents integration over the spot. In the case that lens crosses the position of spot on the lens plane we can estimate the 
time scale of perturbation by $t_{\bullet} \simeq f \times \rho_\star \times t_E$, where $f$ is the ratio of spot size 
to the source size. In the case of binary lensing where the singular area of lens unlike to the single lens is a curved path, the probability of caustic crossing of spot is higher than the case of single lensing \cite{han2000}. 

\subsection{Self-Lensing and observation of compact companions around source star}
Almost two-thirds of stars are in binary systems. For the case that these systems locate at the edge--on position with respect to an observer, the companion star can block light receiving from the other star during eclipsing and result is decreasing the overall light from the two unresolved stars. This method has been used for long time to 
identify binary systems. Using the spectroscopic as well as photometric data,  one can find the parameters of edge-on binary systems. 

\begin{figure}[h]
\centerline{\psfig{file=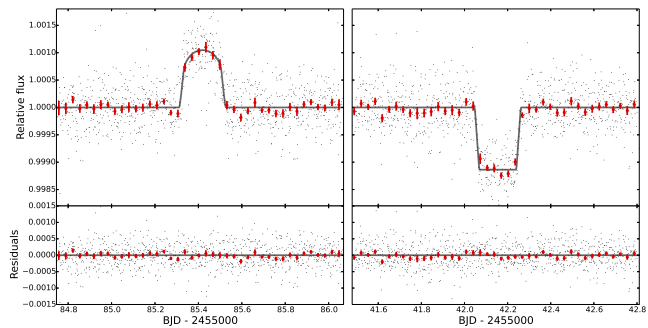,width=10.0cm}}
\vspace*{1pt}
\caption{ Light curve of self-lensing form Kepler Photometry KOI-3278. The left panel is the microlensing 
signal during the self-lensing and right panel is the occultation of white dwarf. The residual between the model and data also is written in the lower panel. The binary system is made of sun-like star and a white dwarf. This figure is adapted form Ref \cite{kepler}.
}
\label{self}
\end{figure}

After evolving one of companions in a binary system, it can turn to a black hole, neutron star or white dwarf. In this case, instead of blocking light of companion star during eclipsing, the compact companion star can magnify light receiving from the other star. Then, we would expect having a small magnification due to self-lensing in the binary system. The first prediction of this effect has been done by Andre Maeder in 1973, long time before the first detection of microlensing event \cite{maeder}. For the short period binary systems, we would expect to observing repeating self-lensing of companion star by its companion. With developing ground based and space based telescopes, it was predicted to observe this phenomenon in practice \cite{sahu, g1,rah1}. 

The first case of self-lensing event in binary system has been discovered in database of Kepler space--telescope in 2014 \cite{kepler}. This event, so-called KOI-3278 is a binary system composed of sun-like star and a white dwarf orbiting it every $88.18$ days.  The microlensing transit occurs in $5$ hours and the result of lensing is $0.1 \%$ magnification in the flux of companion star. Figure (\ref{self}) shows the first self-lensing candidate of KOI-3278 in Kepler data.

\section{Astrometric and Polarimetric Observation}
\label{astrometry}

In this section we will introduce two alternative methods of astrometric and polarimetric observations of microlensing events. In the astrometric observation, we 
are interested in to detect the position of images or in another word the displacement of source position due to 
dynamics of images during the lensing. The other method is the polarimetric observations of microlensing events for those
source that have polarized signal. These two methods will produce extra-informations and enable us to break completely the degeneracy between the lens parameter. Moreover, from the polarimetric observation we can extract extra information about the nature of source stars. The observation of these two parameters will open a new window to study the physics of stellar atmosphere and application of microlensing in the stellar astrophysics. 

\subsection{Astrometry of microlensing events}
In the previous sections we have discussed about the photometry of gravitational microlensing events where the flux of light receiving from a source star becomes brighter during the transit of source over the Einstein ring. The only physical parameter that we can extract from the photometric 
observation is the Einstein crossing time which depends on the mass, distance and velocity of a lens. In this section 
we discuss on possibility of astrometric observation of microlensing events. 

\begin{figure}[h]
\centerline{\psfig{file=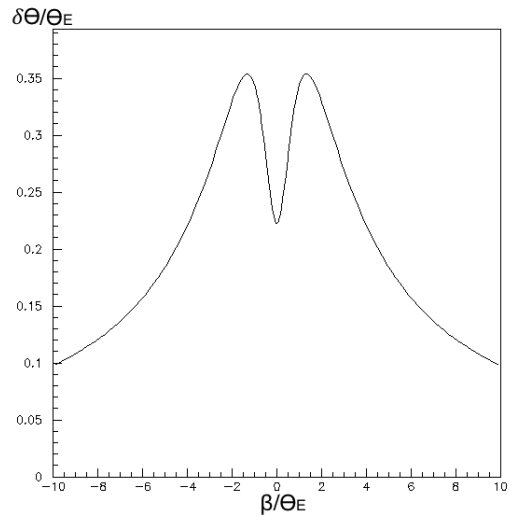,width=7.0cm}}
\vspace*{1pt}
\caption{Astrometric shift of images in microlensing event with the minimum impact parameter of $\beta_0 = 0.5$. The 
impact parameter and astrometric shift angles are normalized to the Einstein angle.}
\label{ast}
\end{figure}

During microlensing by a single lens, two distinctive images produce at the either sides of the lens position. The separation between the images from equation (\ref{sep}) and (\ref{angle}) is in the order of Einstein angle
and it is about $\sim1$~mas. This angular distance between the images is smaller than 
the resolution of ground based telescopes. High resolution space-based telescopes or adaptive optics may resolve 
these images in future\cite{ast1,ast2,ast3}. In the astrometry we can measurement of the displacement of centre of light during the lensing which do not need extremely high resolution to resolve milli-arc second separation between the images from the lensing. Here, we need to identify and follow the centre of Point Spread Function (PFS) of the images during the lensing. The centre of light of images with respect to the position of lens is given by 
\begin{equation}
{\bf \theta_c} = \frac{|A^+|{\bf \theta^+} + |A^-|{\bf \theta^-}}{|A^+| + |A^-|},
\end{equation}
where $|A^+|$ and $|A^-|$ as discussed before are the absolute magnification of two images at 
the angular positions of $\theta^+$ and $\theta^-$.
Substituting the magnification and position of images from equation (\ref{sol1}) results in 
\begin{equation}
{\bf \theta_c} = {\bf \beta}\frac{\beta^2 + 3\theta_E^2}{\beta^2 + 2\theta_E^2}.
\end{equation}
The observable parameter in the gravitational microlensing is the displacement of the centre of light of
source star during the lensing. This parameter can be obtain by $\delta\theta = \theta_c - \beta$ and 
the result is
\begin{equation}
{\bf \delta\theta} = {\bf\beta}\frac{\theta_E^2}{\beta^2 + 2\theta_E^2}.
\label{ashift}
\end{equation}
Figure (\ref{ast}) represents the absolute value of astrometric shift during microlensing for an event with the minimum impact parameter of $\beta_0 = 0.5 \theta_E$. The maximum displacement of centre of light from $\partial|\delta\theta|/\partial|\beta| = 0$ happens at $\beta = \sqrt{2} \theta_E$ which results in the maximum displacement of centre of light, $\delta\theta_{max} = \theta_E/2^{3/2}$. The 
observation of this quantity directly provides the observation of Einstein angle. 

We note that while the maximum photometric magnification happens at the smallest impact parameter, the maximum astrometric shift happens before the source star enters the Einstein ring. The other property of astrometric shift is that at the large impact parameters,  $\delta\theta$ falls as $1/\beta$ while for the magnification $\delta A \propto 1/\beta^4$, means that we can 
detect the astrometric shift even at the larger impact parameters. 
This property results in a larger astrometric optical depth compare to the photometric optical depth \cite{dom,balk}. For the case of GAIA satellite where the accuracy in the astrometric shift is in the order of micro-arc second, the optical depth can be obtain with the 
accuracy limit of telescope. 
In this case from (\ref{ashift}), assuming $\delta_T$ as threshold for the observation, the corresponding maximum 
impact parameter obtain as $\beta = \theta_E^2/\delta_T$. Comparing the astrometric optical depth with the photometric 
optical depth \cite{dom} shows that in the former case optical depth is about $10^{-5}$ while for the later one, it is in the order of 
$10^{-7}$.




In the case of accompanying the astrometric observation with the photometric observations, having high cadence and high photometric precision, it would be 
possible to detect the perturbation effects such as parallax or finite source size effects. This will help us to 
break completely the degeneracy between the lens parameters \cite{rahghas}. Let us assume accompanying astrometric observation with the parallax measurements in the light curve from the ground based of space based observations. In this case we 
obtain the parameter of $\delta u$ which is given in terms of Einstein angle and distance of lens and source as follows $$\delta u = \frac{a_\oplus(1-x)}{D_S x \theta_E}.$$  For the microlensing events in the direction of Galactic Bulge the distance to the source stars is almost $8$~kpc. On the other hand from astrometric observation, we can measure $\theta_E$. Substituting these parameters in the parallax equation, we can extract $x$ parameter which is the relative distance of lens to the source. On the other hand, using the equation for the Einstein angle and distance of lens and source, we can obtain the mass of lens and subsequently the transverse velocity of lens-source from measuring the Einstein crossing time.

The other helpful observation is the finite size effect of source star which enables us to measure the projected radius of source star on the lens plane, normalized to the Einstein 
radius (i.e. $\rho^\star = x R_s/R_E$). The size of source star, after implementing the extinction correction of interstellar medium, can be obtained from the position of star in the colour-magnitude diagram \cite{yoo}. One of the problems 
with identifying the position of the star in the colour-magnitude diagram is the blending effect of either lens star or
stars in the background that can alter the position of star.  

Measuring $\rho^\star$ and $R_s$ can put constrain on $\rho^\star = {R_s} x/{R_E}$ and the result is constraining 
between the mass of lens and distance from the observer. On the other hand from the astrometry we can measure the Einstein angle $\theta_E$ which combining with the finite size effect, we can measure the distance of source star from the 
observer. 
 
\subsection{Polarimetry observation of microlensing events}
The scattering of photons by the atmosphere of stars produces linear polarization. The mechanism of producing polarized light is due to scattering of light by free electrons of atmosphere in the hot stars \cite{chandra}, scattering of light by atoms and molecules in the main sequence stars \cite{Fluri} and scattering by the dust grains in the cold stars \cite{Si}.  The effect of polarization in gravitational microlensing has been studied in Supernova explosions where 
local polarization on the gas of a symmetric explosion can be magnified by another intervening compact object as a gravitational lens \cite{schni}. The application of this phenomenon to the microlensing events within Milky Way galaxy \cite{simon} is aiming to  enhance polarization signals from the source stars during lensing. This method enables observers to study the atmosphere of 
stars thousands of parsec far from Earth, somewhere at the Galactic centre or Spiral arms of Galaxy. 

Assume a symmetric polarization of source star due to the scattering of light by the atmosphere of a star, in the gravitational microlensing, different parts of source disk magnify with different amount and for the smaller impact parameters, the higher is the magnification. The result is breaking symmetry of polarization pattern and producing a net polarization signal from the source star. The extra information from the amount of polarization and time dependence of it during lensing can break the degeneracy between the lens parameters as well as provide a unique information about the atmosphere of source stars \cite{sajadian2014,Jetzer2015}.

\begin{figure}[h]
\centerline{\psfig{file=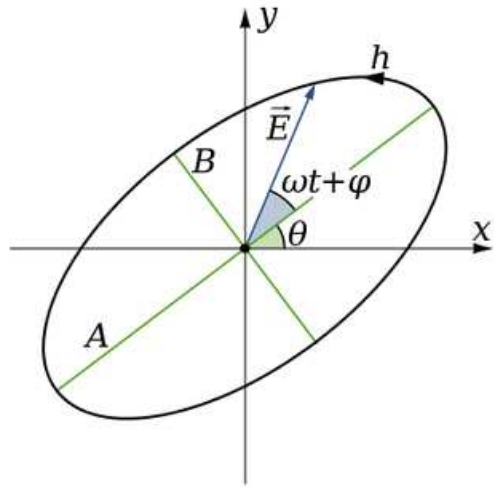,width=7.0cm,angle=0}}
\vspace*{1pt}
\caption{Polarization ellipse with semi-major axis of $A$ and semi-minor axis of $B$. The semi-major axis is rotated with the angle of $\theta$ in counterclockwise direction with respect to the coordinate system. This Figure is adapted from reference \cite{polarization}.}
\label{pol}
\end{figure}

\begin{figure}[h]
\centerline{\psfig{file=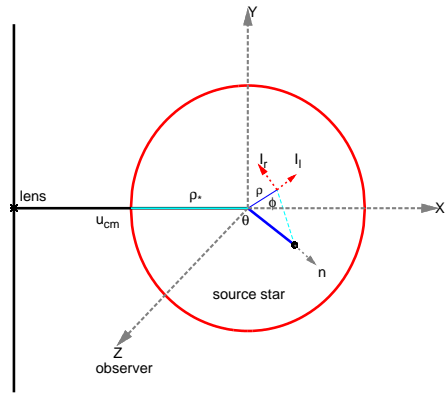,width=7.0cm,angle=-90}}
\vspace*{1pt}
\caption{Projection of source star on the lens plane where $Z$ direction represents the line of sight from the 
observer to the source. The intensity of lens is decomposed into radio direction (i.e. $I_l$) and transverse direction 
(i.e. $I_r$). Here we adapt the convention by Chandrasekhar \cite{chandra}. This Figure is adapted from Reference \cite{sajadian2014}.}
\label{coordinate}
\end{figure}

The polarization of a source star is given in terms of four Stokes parameters which is expressed by $S_I$ , $S_Q$, $S_U$ and $S_V$. Let us assume electromagnetic filed in the directions of $X$ and $Y$ axis perpendicular to the propagation of light, varies as $E'_X = A \exp(i\omega t)$ and  $E'_Y = B\exp(i\omega t + \pi/2)$. Now we rotate coordinate system along the propagation of light, in clockwise direction by the angle of $\theta$ (as shown in Figure {\ref{pol}) and in the new coordinate system calculate 
the Stokes parameters. While the circular polarization for stars with taking into account the stellar spot is non zero\cite{Illinga,Illingb}, for simplicity we set $S_V = 0$ for stars without spots. The  Stokes parameters on the surface of star with the area of $S$ defined as follows:
\begin{eqnarray}
S_I &=& \int_S <|E_X(s)|^2 + |E_Y(s)|^2> d^2S  = \int \left(A(s)^2 + B^2(s)\right) d^2S \nonumber \\
 &=& \int_S I(\mu) d^2S,
\end{eqnarray}
where $I(\mu)$ represents the intensity of star in polar coordinate on source disk and $\mu$ is the cosine between the line of sight and vector normal to the surface of star. Using the Chandrasekhar convention, for the polar coordinate on the surface of a star, 
the intensity of light in the radial and tangent directions are taken as 
$I_l = A^2$ and $I_r = B^2$, see Figure (\ref{coordinate}).

 The other Stokes parameters are
\begin{eqnarray}
S_Q &=& \int_S <E_X^2(s) - E_Y^2(s)> d^2S =\int \left(A(s)^2 - B^2(s)\right)\cos 2\theta  d^2S \nonumber \\
&=&  - \int I_{-}(\mu) \cos 2\theta d^2S, \\
S_U &=&  - \int_S <E_X^\star(s) E_Y + E_Y^\star(s)E_X(s)> d^2S = - \int \left(A(s)^2 - B^2(s)\right)\sin 2\theta d^2S \nonumber \\
&=& \int I_{-}(\mu) \sin 2\theta d^2S, 
\end{eqnarray}
where $I_{-} = B^2 - A^2 = I_l - I_r$.

The apparent intensity of stars for an observer can change due to limb darkening\cite{chandra}
\begin{equation}
I(\mu) = I_0 (1 - c_1(1-\mu)),
\end{equation}
where $I_0$ and $c_1$ are the parameters for limb darkening function and $\mu$ as described before represents the distance from the centre of source disk (e.g. $\mu =1$ is the centre and $\mu=0$ represents limb of star). The high magnification microlensing events can probe the parameters of limb darkening \cite{choi}.
The polarized component of source stars also can be given by 
\begin{equation}
I_{-}(\mu) = I_0 c_2 (1-\mu),
\end{equation}
where $c_2$ represents local polarization fraction on surface of star. 
The parameters of limb darkening and polarization depends on the physics of atmosphere for a given star \cite{Ing2012}. For the hot stars, using the light scatteting by the free electrons in the atmosphere, Chandrasekhar (1960) calculate the limb darkening and polarization from the surface of this type of stars as follows \cite{bel}
\begin{eqnarray}
I_{+}(\mu)&=&\frac{1+16.035\mu+25.503\mu^2}{1+12.561\mu+0.331\mu^2}, \nonumber \\
I_{-}(\mu) &=& \frac{0.1171+3.3207\mu + 6.1522\mu^2}{1 + 31.416\mu + 74.0112\mu^2}(1-\mu).
\end{eqnarray}
 
 For the case of main sequence stars such as Sun, the polarization is mainly due to the Rayleigh scattering by neutral hydrogen and partially due to Thomson scattering by free electrons. The polarization of light for the case of Sun is a wavelength dependent function and it can be modelled by using the structure of atmosphere of Sun. The best fit to the theoretical model is given by \cite{fo1,fo2}
 \begin{equation}
 I_{-}(\mu) = q_\lambda \frac{1-\mu^2}{(\mu + m_\lambda)(I_\lambda(\mu)/I_\lambda(1))},
 \end{equation}
where parameters with the index of $\lambda$ depends on the wavelength. For the case of cool stars, depending on the atmospheric model and ingredients of dusts, the polarization function can change \cite{simon}. Using the magnification factor in the gravitational microlensing, the Stocks parameters can be written as follows:  
\begin{eqnarray}\label{tsparam}
\left( \begin{array}{c} 
S_{Q}\\
S_{U} \\
S_{I}
\end{array}\right)
&=&\rho^2_{\star}\int_{0}^1\rho~d\rho\int_{-\pi}^{\pi}d\theta  A(u)  
\left( \begin{array}{c} - I_{-}(\mu) \cos 2\theta \\
 I_{-}(\mu) \sin 2\theta \\
I(\mu)
\end{array} \right),
\end{eqnarray}
where $u$ is the impact factor (distance of lens from each element on the source star) and $A(u)$ is the local magnification factor.  

It is supposed that the next generation of microlensing experiments will perform polarimetry as well as astrometry observations beside the photometry observations. This supplementary observations will open a new window for scanning the atmosphere of remote stars and investigate the details of their structures. Moreover, using this method, we can study defects on the surface of stars as stellar spots and even measure the strength of their magnetic field.

\section{Summary} 
\label{conc}
In this article we reviewed the astrophysical applications of single-lens gravitational microlensing events. The first microlensing candidate has been observed almost two decades ago and the aim of microlensing experiments at that time was discovering the Massive Astrophysical Compact Halo Objects (MACHOs) in the galactic halo. However, the suggestion of exploring planets in the binary lenses with a parent star and planet orbiting around it, opened a new window to the microlensing observations and since more than ten years, microlensing experiments are mainly focused on exoplanet detections. With this method more than ten exoplanet at the distance of snow-line from the parent stars have been discovered. Moreover, using large field telescopes and follow-up observations, more than 3000 microlensing event have been discovered each year in the direction of Galactic Bulge.  

Using dedicated space based telescope by means of observing events both from Earth and space, it would be possible to do both parallax measurement and astrometric observations. This kind of next generation microlensing experiments could measure the mass of lenses, as well as distance of lenses from the observer. Since part of lenses are too faint to be observed or they are compact objects as black hole, neutron stars or white dwarfs, it would be possible to produce the mass function of stars and stellar remnants. Also knowing the distance of stars, it is possible to study the structure of the Milky Way and distribution of stars along the line of sight.

The other phase of microlensing observation would be the astrophysical application of this method as a natural telescope to scan the surface of source stars. This application can be done by photometric  and polarimetric channels. With this method one can detect small defects as the stellar spots on the surface of stars even located at the Bulge of Milky galaxy. Regarding vast applications of microlensing, it seems that this technique will be an important astrophysical tool to study the stellar physics as well as the structure of Milky Way galaxy.





\end{document}